\documentclass[twocolumn,english,aps,prb,reprint, superscriptaddress,showpacs,longbibliography,showkeys,nofootinbib]{revtex4-2}
\usepackage{amsmath,amssymb,bbm,mathrsfs,bm,color,graphicx,comment,xcolor,dsfont,multirow}
\usepackage[colorlinks,citecolor=blue,urlcolor=blue,linkcolor = blue]{hyperref}
\usepackage{enumitem}
\usepackage[mathscr]{euscript}
\usepackage[T1]{fontenc}
\usepackage{siunitx}
\usepackage{subfigure}
\usepackage{adjustbox}
\usepackage{booktabs}
\usepackage{tikz}
\usetikzlibrary{quantikz2}

\begin{document}

\title{From NISQ to Fault-Tolerance: Applications and Algorithmic Benchmarks for Spin Qubits}

\author{Frederik Lohof}
    \affiliation{Parity Quantum Computing Germany GmbH, Schauenburgerstraße 6, 20095 Hamburg, Germany}

\author{Florian Ginzel}
    \affiliation{Parity Quantum Computing Germany GmbH, Schauenburgerstraße 6, 20095 Hamburg, Germany}

\author{Wolfgang Lechner}
    \affiliation{Parity Quantum Computing Germany GmbH, Schauenburgerstraße 6, 20095 Hamburg, Germany}
    \affiliation{Parity Quantum Computing GmbH, Rennweg 1, Top 314, 6020 Innsbruck, Austria}
    \affiliation{Institute for Theoretical Physics, University of Innsbruck, 6020 Innsbruck, Austria}

\date{\today}

\begin{abstract}
    Motivated by recent breakthroughs in the development of spin-based quantum processing units based on exchange-only (EO) spin qubits, we provide a roadmap for the implementation of quantum algorithms on the EO platform, ranging from the NISQ to the fault-tolerant era. To provide an algorithm-driven perspective on the scaling of quantum chips, we consider a range of applications targeting different stages of hardware maturity and formulate requirements for a successful realization. We show that the compilation method Parity Twine perfectly complements the hardware's capabilities to perform tasks such as the quantum Fourier transform or QAOA. Furthermore, we describe an error detection technique native to Parity Twine, which EO qubits can leverage in a unique and advantageous way to improve algorithm performance. Finally, since both near-term algorithmic benchmarks and a long-term perspective can be found in digital quantum simulation, we specifically discuss the fermionic fast Fourier transform and the simulation of Fermi-Hubbard models. The latter is explicitly discussed in the context of quantum error correction and a partially fault-tolerant realization. By providing detailed resource estimates and identifying scaling bottlenecks on each level, our work offers a quantitative perspective on EO-based quantum computing and will inform future hardware design choices.
\end{abstract}

\maketitle

\section{Introduction\label{sec_introduction}}

Semiconductor spin qubits are rapidly emerging as a leading platform for quantum computing, positioning the technology for near-term applications while providing a clear pathway toward scalability~\cite{hrl_qpu,Dijkema2026,DumoulinStuyck2026,McIntyre2026,burkard_semiconductor_2023}. A particularly interesting property of spin qubits is the ability to encode quantum information in a decoherence-free subspace (DFS) using multiple exchange-coupled electrons~\cite{PhysRevLett.85.1758,DiVincenzo2000,PhysRevLett.89.147902,PhysRevB.95.241303,Hoffmann2026}. Among these DFS qubits, the exchange-only (EO) qubit, comprising three spins in three quantum dots, stands out as the smallest system where the pairwise exchange coupling allows for universal control~\cite{DiVincenzo2000,PhysRevLett.85.1758}. Therefore, EO qubits allow for all-electrical baseband control without the need for magnetic field gradients~\cite{weinstein_universal_2023}.

All gate operations on EO qubits are composed of sequences of shape-insensitive, sub-microsecond exchange pulses~\cite{DiVincenzo2000,FongWandzura,Russ2017}, making the control versatile, robust and orders of magnitude faster than their long $T_2$ times~\cite{hrl_qpu}. The development of efficient pulse sequences enables a diverse set of encoded single- and two-qubit gates for EO qubits~\cite{Chadwick2025,heinz_fast_2025}, while the potential for parallel pulse execution offers opportunities to accelerate algorithm execution~\cite{madzik_operating_2025}. In future application, their properties may make EO qubits excellent building blocks of fault-tolerant architectures~\cite{Chadwick2025_QEC,Heinz2026}.

Recent breakthroughs have addressed many limiting bottlenecks of spin qubits. Building on previous developments~\cite{Ha_flexible_2022,weinstein_universal_2023,Ha_two_dimensional_2025}, an integrated quantum processing unit (QPU) incorporating up to 18 EO qubits, a digital cryo-CMOS controller and sophisticated interconnects was demonstrated~\cite{hrl_qpu}. While the connectivity of near-term devices is still expected to be limited to quasi-linear chains~\cite{Li2026,Dijkema2026,hrl_qpu}, this represents a major leap in hardware maturity.

Remarkably, the high fidelities reported in Ref.~\cite{hrl_qpu} are limited by the interplay of the various components rather then intrinsic noise sources, thus pushing spin qubits into a new phase of engineering. Optimizing individual components alone is no longer sufficient and further progress will require a holistic view on the QPU and the interactions between its subsystems. Crucial to that view is a mapping of the target algorithm to the hardware layer that takes component interactions within the device into account, and has strong implications for the QPU performance~\cite{Lechner2020,Fellner2022applications,ginzel_scalable_2024,runtime_paper,MontanezBarrera2025,twine_demo,aigner_fermion_2026}. An inefficient mapping, unaware of the QPU's strengths and bottlenecks, can incur an overhead in interactions and circuit depth, increasing the vulnerability to errors and thus limiting the process fidelity.

Recently, Parity Twine has emerged as a powerful method for implementing certain classes of quantum algorithms on QPUs in a way that is aware of the hardware's connectivity constrains~\cite{twine}. Notably, it was used to demonstrate a record-breaking process fidelity for the quantum Fourier transform (QFT) using 50 superconducting qubits~\cite{twine_demo} and it was shown to offer the most efficient approach to quantum optimization to date~\cite{MontanezBarrera2025}. Another recent discovery is a method for the simulation of fermion lattices using qubit lattices of the same size with $\mathcal O (1)$ interaction overhead, even improving the asymptotic scaling of certain tasks in materials science and quantum chemistry~\cite{aigner_fermion_2026}.

Such circuits can be implemented directly on physical qubits for noisy, intermediate-scale quantum computing (NISQ). However, even with the most efficient compilation schemes the accumulation of errors inevitably makes quantum error correction (QEC) and fault-tolerant quantum computing necessary as the computational tasks grow~\cite{Preskill1998,NielsenChuang,Preskill2018quantumcomputingin}. QEC adds an additional level of encoding where repeated measurements of stabilizers of the code allow for the detection and suppression of errors~\cite{PhysRevA.109.032433}. Depending on the physical noise level, QEC introduces a significant resource overhead, effectively demanding to balance qubit engineering, fabrication and control capabilities, and algorithmic requirements~\cite{PRXQuantum.5.040328,Steinacker_2026_temperature}.

\begin{figure}[htb]
    \centering 
    \includegraphics[width=.99\columnwidth]{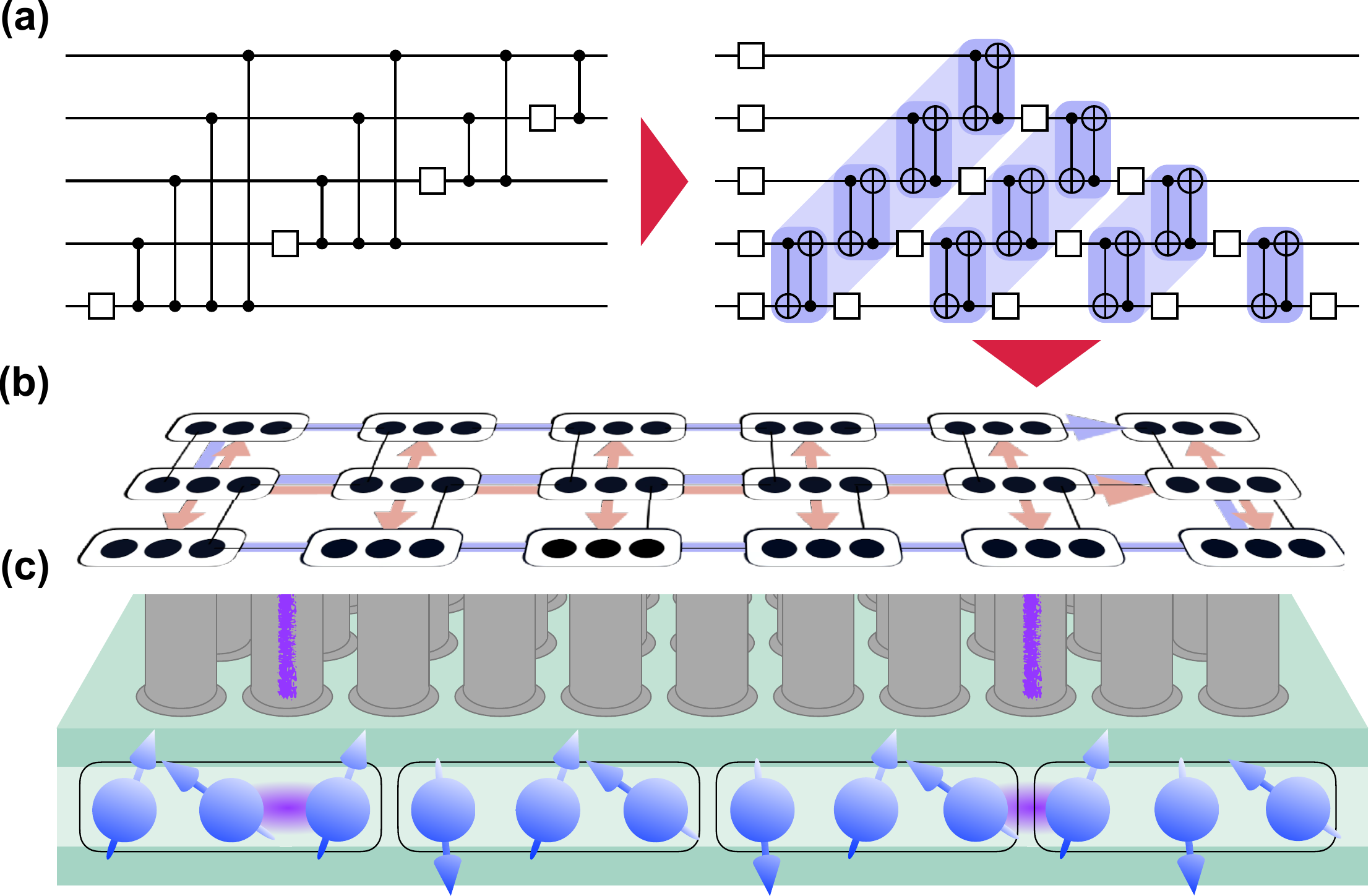}
    \caption{Compilation and execution pipeline of a quantum algorithm on a EO-based QPU as studied in this paper. (a) A general quantum algorithm like the QFT is compiled to respect the connectivity of the quantum chip. We choose the Parity Twine compilation which conveys parity information along chains of double $\mathrm{CX}$ ($\mathrm{DCX}$) gates (purple). (b) The Parity Twine network must be mapped to the quantum chip. Generally, this mapping is not unique and different options, e.g., a linear chain in a boustrophedon patter (purple) or a fishbone structure adapted to square-grid connectivity (salmon) must be weighed against each other. Each EO qubit (black boxes) comprises three physical spins (black dots). The connectivity between the qubits (black lines) and the spins determines the pulse sequences for logical gates. (c) On the physical level, logical operations correspond to sequences of pulses of exchange interaction (violet) between neighboring spin-$1/2$ particles (blue) which are confined in a semiconductor heterostructure (teal). The interaction is controlled by DC voltages applied to electronic contacts (silver), crosstalk determines the shortest distance between active couplings and may thus limit parallelization.
    All layers are interlocked and the performance of the algorithm is critically dependent on their interactions and synergies.}
    \label{fig_overview}
\end{figure}

In this paper we aim to chart a path from first proof-of-principle demonstrations and algorithmic benchmarks to practical applications of a QPU based on EO qubits. As illustrated in Fig.~\ref{fig_overview}, we compile important quantum algorithms to EO qubits and optimize further at the level of elementary exchange pulses. With this scheme we can directly harness the powers of the EO platform. Our resource estimates provide guidance for optimally realizing applications appropriate at different stages of hardware maturity.

For the near term, EO qubits exhibit an ideal synergy with Parity Twine, which relies on chains of composite $\mathrm{CX} \odot \mathrm{SWAP}$ gates, also referred to as $\mathrm{DCX}$ gates. Leveraging a short pulse sequence for $\mathrm{DCX}$ gates~\cite{Chadwick2025}, we provide resource estimates for the Quantum Fourier Transform (QFT) and Quantum Approximate Optimization Algorithm (QAOA) with Parity Twine and the previously best-performing compilation method. We further present and study an error-detection protocol which is native to Parity Twine and uniquely well-suited for EO qubits. This allows for the mitigation of errors once a moderate overhead in qubit numbers is acceptable. Finally, as a bridge to utility-scale quantum computing, we discuss applications related to quantum chemistry. Specifically, we formulate requirements and target values for the fermionic fast Fourier transform and a partially fault-tolerant simulation of a Fermi-Hubbard model.

This paper is organized as follows. In Sec.~\ref{near_term} we study the performance of Parity Twine-based algorithms on specific chip topologies of EO qubits. After identifying the optimal implementation, we derive resource estimates for the QFT and QAOA. Subsequently, in Sec~\ref{sec_error_detection}, we complement the algorithmic performance with an error detection method. In Sec.~\ref{sec_fermion_appl} we present resource estimates for the simulation of fermion lattices with physical and logical qubits. Finally, in Sec.~\ref{sec_summary}, we summarize our results and provide an outlook.

\begin{figure}[t]
    \centering
    \includegraphics[width=.95\columnwidth]{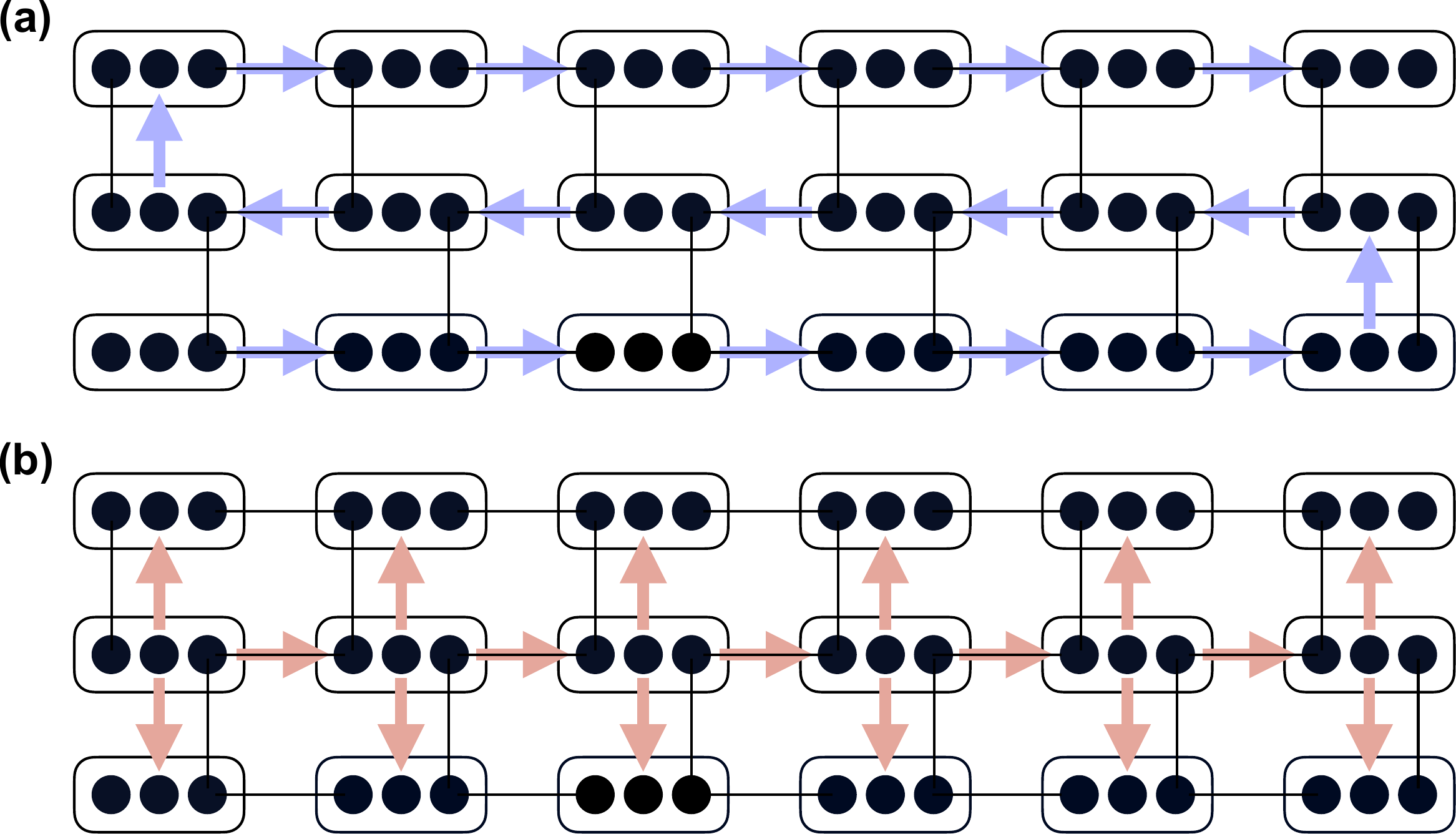}
    \caption{Schematic PTN layouts mapped to a $3\times L$ EO qubit device, with $L=6$ as example. Black lines indicate physical spin connections, while arrows indicate the orientation of $\mathrm{DCX}$/$\mathrm{CX}$ gates that distribute parity labels on the qubits~\cite{twine}. (a) The linear-chain PTN is defined by chains of $\mathrm{DCX}$ gates winding from row to row (blue arrows) (b) The square-grid PTN defines the central row as a backbone $\mathrm{DCX}$ chain that is extended by vertical $\mathrm{CX}$ connections (red arrows).}
    \label{fig:PTN_layouts}
\end{figure}

\section{Near-Term Quantum Algorithms with Exchange-Only Qubits\label{near_term}}

In this section we compile near-term quantum algorithms based on the Parity Twine method on state-of-the-art EO chip architectures and provide resource estimates that allow to assess the feasibility of algorithms as function of problem size. To account for the physical reality of EO qubits, metrics like $\mathrm{CX}$ gate counts and circuit depth on this layer are not sufficient. The wide range of known single- and two-qubit gate pulse sequences for EO qubits as well as the complexity of the EO-internal spin connectivity require a compilation approach specific to the EO hardware and a resource analysis on the hardware pulse level.

Specifically we investigate the quantum Fourier transform (QFT) as well as the quantum approximate optimization algorithm (QAOA). Doing this we utilize the existence of exchange pulse sequences for EO qubits~\cite{Chadwick2025} that are especially beneficial to the Parity Twine method. We adopt the chip layout and connectivity map from Ref.~\cite{hrl_qpu} for near-term applications, an array of $3\times 3L$ quantum dots forming $3\times L$ EO qubits, Fig.~\ref{fig_overview}(b), due to its relatively relaxed fabrication requirements~\cite{Li2026}. Our results demonstrate that the Parity Twine approach provides distinct advantages, yielding a significant reduction in resource overhead compared to previous state-of-the-art compilation methods.

\begin{figure*}[htbp]
    \centering
    \includegraphics{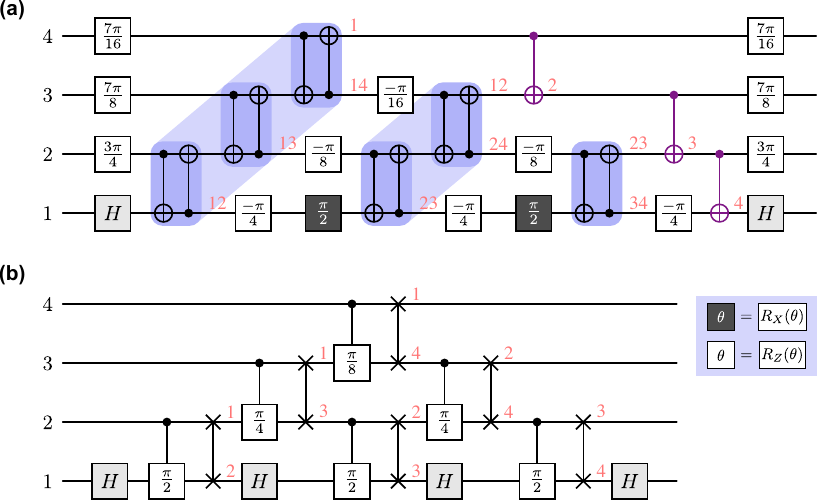}

    \caption{(a) Circuit for linear-chain Parity Twine QFT for $N=4$ qubits. Blocks of two $\mathrm{CX}$ gates (blue boxes) are implemented as a single $\mathrm{DCX}$ on the EO qubit platform. Chains of $\mathrm{DCX}$ gates (shaded areas) generate all possible two-body parity labels (red numbers). Colored $\mathrm{CX}$ belong to the decoding step restoring the original single-particle labels. (b) Reference QFT implementation with linear connectivity introduced by Fowler \textit{et al.}~\cite{fowler_QFT}. The circuit implements a SWAP network to mediate the all-to-all connectivity required by the standard QFT.}
    \label{fig:QFT_circuits}
\end{figure*}

Parity Twine utilizes parity labels to track the flow of information through the qubit array~\cite{flow}. A qubit with label $ij$ encodes the parity of base qubits $i$ and $j$, ${Z}_{ij} = {Z}_i {Z}_j$. While encoded in parity qubits, the quantum state permits the implementation of $e^{i\alpha Z_i Z_j} = e^{i\alpha Z_{ij}}$ as single-qubit phase gates. The label $ij$ can be obtained as the result of a $\mathrm{CX}$ gate between qubits $i$ and $j$, while a double $\mathrm{CX}$ ($\mathrm{DCX} = \mathrm{CX} \odot \mathrm{SWAP}$) creates a new label and propagates the control label to the target qubit. This is the building block of a Parity Twine network as illustrated in Fig.~\ref{fig:QFT_circuits}(a), where Parity Twine chains of $\mathrm{DCX}$ gates create trails of new labels and single-qubit gates operate on the encoded state. The structure of the network can efficiently be tailored to any connectivity by introducing a backbone of $\mathrm{DCX}$ and legs of $\mathrm{CX}$ gates~\cite{twine}. The correct sequence of single-qubit $Z$- and $X$-rotations to be interleaved with the Twine chains is obtained from the Parity Twine compilation~\cite{swapless,twine}.

In order to turn the Parity Twine circuit for a given algorithms into pulse counts and pulse-circuit depths we rely on optimal pulse sequences for individual two-qubit gates~\cite{Chadwick2025}. Since EO qubits consist of three electron spins in three quantum dots, there are six equivalent configurations encoding the same qubit and the sequence length of two-qubit gates depends on the initial configurations of both qubit's internal spins. If the gate is allowed to permute the final configuration in each EO qubit, even shorter sequences become possible. We use a custom circuit-to-pulse compilation tool to map the circuit acting on logical EO qubits to the pulse circuits acting on individual spins, while respecting the dot connectivity of the underlying chip architecture. It tracks the internal spin configuration of each EO qubit during the circuit and selects appropriate pulse sequences from a database. We execute each pulse as early as possible and also allow for permutations of intra-EO spin positions to reduce the overall pulse count.

The depth of the resulting pulse sequence is strongly dependent on the hardware's capability to perform parallel pulse operation. While fully parallel operation has been demonstrated~\cite{madzik_operating_2025} it relies on the availability of an extensive, well-calibrated instruction set that accounts for the possible combinations of simultaneously active pairs. Sequential pulses result in a smaller and simpler and instruction set, which may be favorable when accounting for limitations of the control hardware. We model this limitation by introducing an average parallel-operation exclusion radius $r$. It represents the separation between two spins, measured in Manhattan distance in dots,  that excludes them from simultaneous operation. To determine the pulse-circuit depth resulting from varying exclusion radii, we use an heuristic critical-path algorithm for pulse scheduling. See Appendix~\ref{sec:appendix} for benchmarks of the scheduling algorithm.

\subsection{Parity Twine Network Resource Estimation}

A Parity Twine Network (PTN) serves as the primary building block for any Parity Twine algorithm \cite{twine}. This network of $\mathrm{DCX}$ and $\mathrm{CX}$ gates efficiently generates all possible two-body parity labels at a given stage within the circuit. First, we focus our analysis on PTNs because they form the foundational core of Twine-based implementations, and because they consist exclusively of two-qubit gates, which dominate the overall exchange pulse count compared to single-qubit gates. Parity Twine can be implemented in various ways, adapted to the connectivity of the underlying qubit system. A EO-chip layout with $3\times L$ qubits naturally supports both a linear-chain PTN and a layout adapted for square-grid connectivity. A schematic of the investigated chip topology and the mapping of the two PTN layouts onto the architecture is shown in Fig.~\ref{fig:PTN_layouts}. The selection of the PTN layout determines the relative orientation of EO qubits connected by two-qubit gates in the network. We start by investigating the effect of this choice in order to determine the layout that is more beneficial on the investigated device topology. 

First, we analyze the two-qubit gate counts for these distinct PTN layouts and convert them into pulse counts averaged over all intra-EO spin permutations. The linear-chain PTN, Fig.~\ref{fig:PTN_layouts}(a), is constructed from $\mathrm{DCX}$ chains that are mapped onto the chip, row by row, in a boustrophedon pattern. Additionally, a single chain of $\mathrm{CX}$ gates traversing the path in the opposite direction is used to decode the original one-body labels. This configuration results in two-qubit gate counts of $N(N-1)/2$ $\mathrm{DCX}$ and $N-1$ $\mathrm{CX}$ gates. In contrast, the square-grid PTN utilizes a more complex construction that leverages the increased connectivity of the square grid relative to a linear topology~\cite{twine}. The corresponding gate counts are $N(N+1)/6-1$ $\mathrm{DCX}$ and $N(N+1)/3$ $\mathrm{CX}$ gates.

\begin{figure}[t]
    \centering 
    \includegraphics[width=.99\columnwidth]{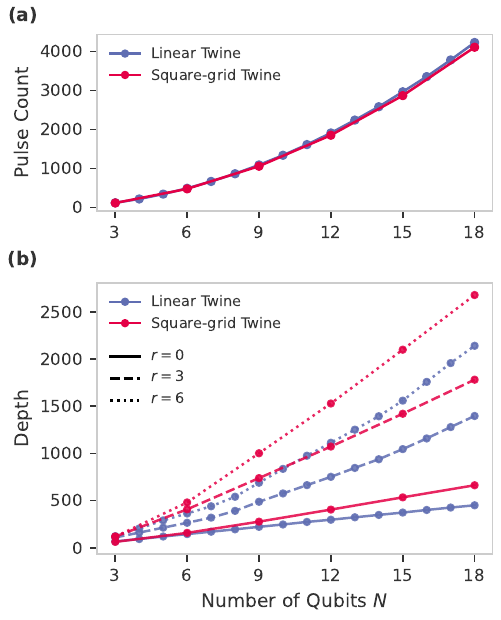}
    \caption{Resource estimates for PTNs at fixed device length $L=6$. (a) Pulse count for linear (blue) and square-grid (red) connectivity as a function of EO qubit number $N$. Twine-chain layouts follow those in Fig.~\ref{fig:PTN_layouts}. Both connectivity types exhibit quadratic scaling with similar leading-order coefficients. (b) Circuit depth measured in exchange pulses for both PTN types as a function of $N$. Line styles represent varying parallel exclusion radii $r$. For Linear Twine, a finite exclusion radius increases the depth scaling after every turn of the linear chain in the chip.} 
    \label{fig:PTN_analysis_fixed_length}
\end{figure}

If $\mathrm{DCX}$ gates must be decomposed into two $\mathrm{CX}$ gates, the square-grid connectivity provides an advantage by utilizing a larger proportion of single $\mathrm{CX}$ gates to distribute parity information among qubits. The encoded $\mathrm{DCX}$ gate provided by EO qubits negates this advantage and results in the same total number of two-qubit gates for both implementations. Fig.~\ref{fig:PTN_analysis_fixed_length}(a) shows the average pulse count for both PTN implementations as a function of the number of qubits for a fixed device length $L=6$. Both Twine layouts scale similarly. Given that average pulse counts are 23.33 for $\mathrm{CX}$ and 25.67 for $\mathrm{DCX}$ gates~\cite{Chadwick2025}, the square grid shows only a slight reduction in pulse count due to the larger proportion of $\mathrm{CX}$ gates involved. 

\begin{figure}[t]
    \centering 
    \includegraphics[width=.99\columnwidth]{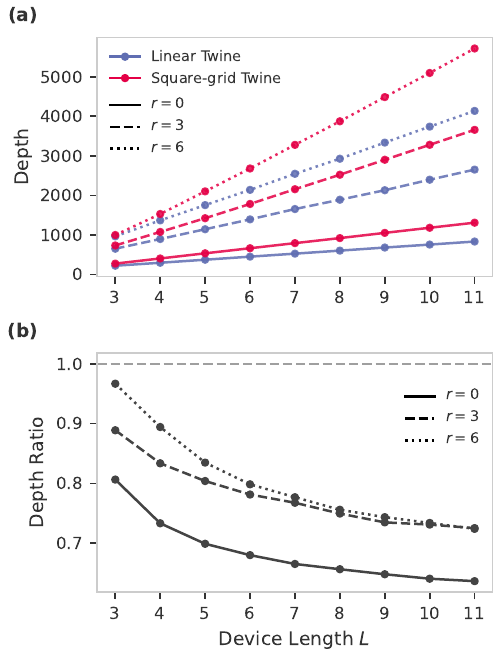}
    \caption{Resource estimates for PTNs at variable device length $L$, resulting in total qubit number $N=3L$. (a) Circuit depth measured in exchange pulses for both connectivity types as a function of $L$. Line styles represent varying parallel exclusion radii. (b) Depth ratio linear-chain/square-grid Twine. The EO qubit connectivity utilized in linear-chain Twine enables greater parallelization of pulses due to the regular relative orientation of spins along the 1D chain.} 
    \label{fig:PTN_analysis_variable_length}
\end{figure} 

\newcommand{\boxl}[1]{\gate[style={inner sep=-0.1cm, fill=white}][0.8cm][0.8cm]{#1}}
\newcommand{\boxd}[1]{\gate[style={inner sep=-0.1cm, fill=black!70}, label style={color=white}][0.8cm][0.8cm]{#1}}
\newcommand{\boxh}{\gate[style={inner sep=-0.1cm, fill=black!10}][0.8cm][0.8cm]{H}}

The error rate of any algorithm implementation does not only depend on pulse count, via the accumulation of gate errors, but also significantly on algorithm depth, due to dephasing errors. Therefore, we investigate the depth reduction that can be achieved by parallel pulse operation in the compiled circuit. Applying exchange pulses in parallel requires complex instructions~\cite{madzik_operating_2025} to reduce crosstalk, an effort that compounds as the system size increases. Fig.~\ref{fig:PTN_analysis_fixed_length}(b) illustrates the pulse circuit depth after scheduling pulses, comparing the two PTN layouts for increasing exclusion radii $r$. If fully parallel operation ($r=0$) can be achieved, the depth scales linearly for both layouts. A finite exclusion radius ($r>0$) increases the depth scaling. For linear-chain Twine the reversal of $\mathrm{DCX}$-chain directions at the end of each row leads to a visible increase in depth scaling.

To asses which PTN layout is more beneficial on the investigated EO-chip topology, we determine the depth as a function of the total device length $L$, with total EO-qubit number $N=3 L$, assuming utilization of all qubits on the device. Results are shown in Fig.~\ref{fig:PTN_analysis_variable_length}(a). The overall linear scaling behavior matches the one predicted by Parity Twine~\cite{twine} while the scaling prefactor strongly depends on the exclusion radius $r$. Fig.~\ref{fig:PTN_analysis_variable_length}(b) shows the depth ratio of linear-chain vs. square-grid PTNs. Even for large exclusion radius ($r=6$), the linear-chain PTN offers a smaller depth asymptotically approaching over 20\% depth reduction compared to the square-grid layout. Based on these results, in the following sections we focus on linear-chain Twine algorithms.

\subsection{Quantum Fourier Transform}
\begin{figure}[t]
    \centering
    \includegraphics[width=.99\columnwidth]{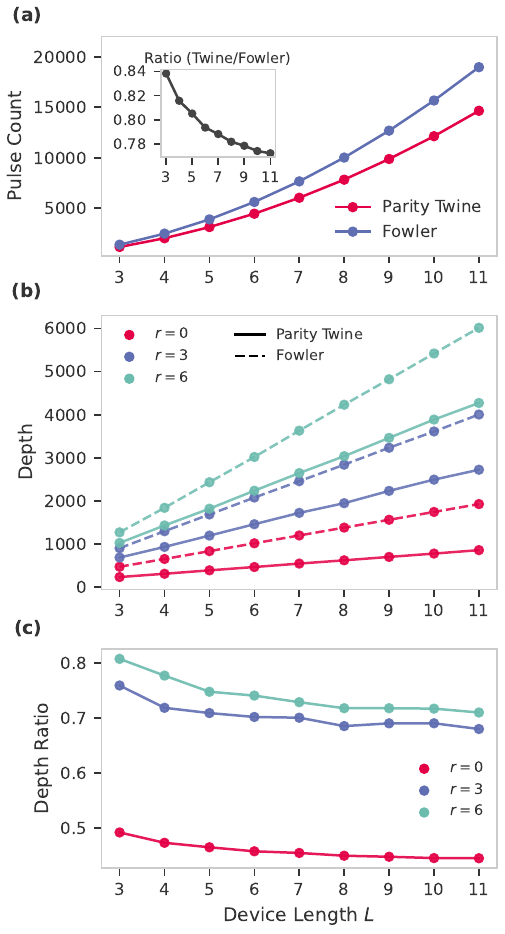}
    \caption{QFT resource estimate. (a) Pulse count scaling comparing linear Parity Twine QFT (red) with the reference Fowler QFT implementation (blue) as a function of device length $L$. Both implementations scale quadratically. The inset shows the pulse count ratio of Twine/Fowler with more than 22\% asymptotic reduction in pulse count for Parity Twine QFT. (b) Circuit depth scaling comparing both QFT implementations for increasing exclusion radii $r$. (c) Depth ratio of Twine/Fowler indicating more than 50\% reduction for parallel pulse operation while the reduction is less pronounced for finite exclusion radii.} 
    \label{fig:QFT_resources}
\end{figure}

The quantum Fourier transform is amenable to implementation using the Parity Twine method. We provide resource estimates for a full QFT on the investigated EO chip architecture and compare it to the previously best-performing implementation by Fowler \textit{et al.}~\cite{fowler_QFT}. It is worth noting that while earlier conjectures suggested that a linear-chain SWAP network, as used by Fowler, was the optimal limit even for square-grid connectivity, the Parity Twine method has refuted this, showing that increased connectivity can be successfully leveraged to reduce resource requirements~\cite{twine}. Despite this potential for higher-degree connectivity, we focus our comparison on the linear-chain Twine approach, as our previous findings indicate it is most beneficial for the specific device architecture under consideration. Example QFT circuits for both, the linear-chain Parity Twine and for the Fowler implementation, on $N=4$ qubits are shown in Fig.~\ref{fig:QFT_circuits}.

Our analysis of the Parity Twine QFT is based on the $\mathrm{DCX}$ and $\mathrm{CX}$ pulse sequences described in Ref.~\cite{Chadwick2025}. Single-qubit operations are implemented as a single-pulse sequence for pure phase gates $R_Z(\theta)$ and a three-pulse sequence for all other single-qubit rotations~\cite{weinstein_universal_2023}. The reference QFT implementation, illustrated in Fig.~\ref{fig:QFT_circuits}(b), constitutes a linear-chain SWAP network that permutes the logical qubit positions to allow for controlled phase ($\mathrm{CP}$) operations only on neighboring qubits. The theoretical gate counts presented in Ref.~\cite{twine} predict an asymptotic two-qubit gate count reduction by a factor of $\frac{2}{3}$ for linear-chain Twine over this reference implementation and a circuit depth reduction by the same factor. However, these results are based on the assumption that all $\mathrm{CP}$ and $\mathrm{SWAP}$ gates in the circuit are decomposed into $\mathrm{CX}$ gates and single-qubit rotations. These results are not valid on EO-qubit-based platforms which offer a wide range of different two-qubit gate implementations.

In this particular case we can make use of a known pulse sequence for the parameterized $\mathrm{CP}$ gate~\cite{zeuch_efficient_2020} and efficient $\mathrm{SWAP}$ gates~\cite{Chadwick2025}. This results in a shorter overall exchange pulse sequence than a decomposition into single-qubit rotations and fixed-angle two-qubit gates. The circuit for Parity Twine QFT matches exactly one $\mathrm{DCX}$ gate with one $\mathrm{CP}$+$\mathrm{SWAP}$ combination. From considering the average pulse counts for the involved gates (25.67 for $\mathrm{DCX}$, 34 for $\mathrm{CP}$+$\mathrm{SWAP}$) we expect an asymptotic pulse count ratio of $\sim 0.75$ between the two implementations. 

In Fig.~\ref{fig:QFT_resources}(a), the pulse count scaling of the Parity Twine QFT (red) is compared to the Fowler reference implementation (blue) as a function of the device length $L$. Although both implementations follow a quadratic scaling, as is theoretically predicted, the linear-chain Twine demonstrates a significant reduction in resource requirements. The inset depicts the pulse count ratio (Twine/Fowler), revealing that while non-leading-order effects are significant for smaller systems, the Twine implementation asymptotically reduces the pulse count by more than 22\%. The apparent mismatch to the predicted 25\% reduction mentioned above is due to the influence of single-qubit gates which are more numerous in the Parity Twine QFT circuit. 

Fig.~\ref{fig:QFT_resources}(b) illustrates the circuit depth for both implementations under different exclusion radii $r$ for parallel gate execution. In all cases, depth scales linearly with system size, with the Twine architecture outperforming the reference implementation. The depth ratio presented in Fig.~\ref{fig:QFT_resources}(c) indicates an asymptotic depth reduction of at least 25\% for exclusion exclusion radii up to $r=6$ and more than 50\% reduction over the reference QFT reference implementation is possible if fully parallel gate operation. These findings highlights the unique suitability of the EO qubit platform for the implementation of Parity-Twine-based QFT to demonstrate near term applications of EO qubits. 

We note that as the problem size increases, the QFT requires exponentially smaller angles~\cite{fowler_QFT}. For current error rates we estimate that with around 16 qubits the smallest rotations are comparable to noise floor, leading to a further degradation of fidelity for larger $N$~\cite{hrl_qpu}. While repetition and post-processing can push beyond this limit~\cite{twine_demo}, eventually the approximate QFT may be a more viable option for scaling up~\cite{PhysRevA.54.139}.

\subsection{Quantum Approximate Optimization Algorithm}

The quantum approximate optimization algorithm, QAOA, is another potential application of near-term quantum hardware that can be efficiently implemented using the Parity Twine method~\cite{swapless, twine}. Its a quantum-classical variational algorithm for solving binary optimization problems encoded in the ground state of Ising-like problem Hamiltonians. Given the problem Hamiltonian 
\begin{equation}
    H_P = \sum_{j=1}^n \sum_{k<j}J_{jk}Z_jZ_k + \sum_{j=1}^n h_j Z_j \label{eq:cost_hamiltonians}
\end{equation}
 the QAOA algorithm applies parameterized circuit layers to a trial initial state, realizing alternating cost and driving unitaries, $U_P(\beta_i)=\exp(-i\beta_i H_P)$ and $U_X(\alpha_i)=\exp(-i\alpha_i \prod_{j=1}^n X_j)$, where $\alpha_i, \beta_i$ are variational parameters and $i\in \{1,\ldots,p\}$. A classical optimization loop determines parameters that maximize the probability of sampling the Hamiltonian's ground state when measuring at the end of the optimized circuit. 

\begin{figure}[tbp]
    \centering
    \includegraphics{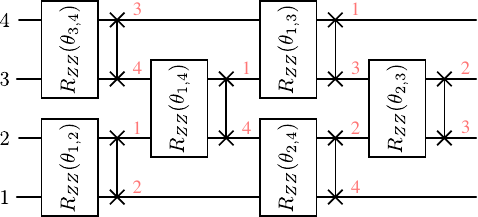}

    \caption{SWAP-network implementation of the QAOA cost layer for $N=4$ qubits. $R_{ZZ}$ gates implement the interactions terms of the cost Hamiltonian in Eq.~\eqref{eq:cost_hamiltonians} with angles $\theta_{j,k} = \beta J_{j,k}$. Red numbers indicate the logical qubit position after each SWAP layer.}
    \label{fig:QAOA_swap_network}
\end{figure}

Since all interaction terms $R_{ZZ}(\theta) = \exp(-i\frac{\theta}{2}Z_jZ_k)$ in each layer commute, Parity Twine can implement them by inserting single-qubit phase gates $R_Z$ into the PTN at the corresponding parity labels. The previously-best implementation, used as a reference, is a SWAP network on a linear chain~\cite{crooks_performance_2018, weidenfeller_scaling_2022}. Even as the investigated the chip topology offers a square-grid connectivity, implementing a SWAP network on the square-grid directly does not provide a benefit over projecting a linear path onto the chip~\cite{weidenfeller_scaling_2022}. The reference SWAP network for a linear chain is shown in Fig.~\ref{fig:QAOA_swap_network}. The circuit structure is similar to the QFT implementation in Fig.~\ref{fig:QFT_circuits}(b), given that the $R_{ZZ}(\theta)$ rotation gates are locally equivalent to $\mathrm{CP}$ gates with adjusted angle arguments. While in the QFT circuit the concrete order of applied $\mathrm{CP}$ operations is determined by the algorithm, the commuting interaction terms in QAOA allow for an implementation with reduced circuit depth in comparison to QFT. 

\begin{figure}[t]
    \centering
    \includegraphics[width=.99\columnwidth]{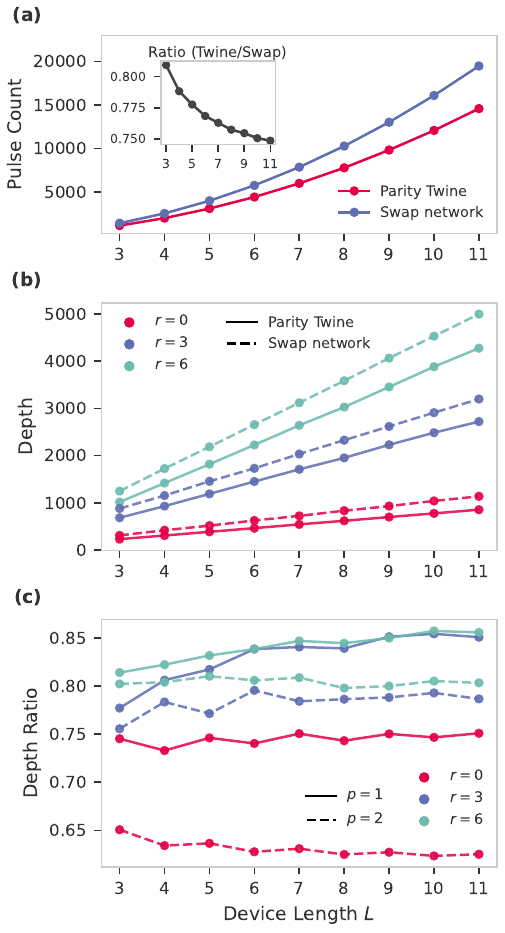}
    \caption{Single-layer ($p=1$) QAOA resource estimates. (a) Pulse count scaling for QAOA as function of device length $L$ comparing Parity Twine QAOA with the implementation using SWAP networks. (b) Depth scaling for increasing exclusion radii $r$. (c) Depth ratio between the two implementation strategies as function of device length $L$. Parity Twine QAOA consistently outperforms the reference implementation using SWAP networks, even as the depth advantage reduces for increasing the exclusion radii $r$.} 
    \label{fig:QAOA_resources_1}
\end{figure}

Fig.~\ref{fig:QAOA_resources_1}(a) compares the pulse count of Parity Twine QAOA for a single layer ($p=1$) with the implementation based on a SWAP network as function of the device length $L$. While both implementations scale quadratically, Parity Twine QAOA asymptotically uses over 25\% fewer pulses as seen in the inset. The depth scaling of the single-layer QAOA circuit is shown in Fig.~\ref{fig:QAOA_resources_1}(b) for varying exclusion radii $r$. The scaling becomes asymptotically linear with prefactors strongly dependent on the exclusion radius $r$. Panel (c) shows the corresponding depth ratio between the two implementations one and two QAOA layers. Notably, the theoretical depth scaling for a $\mathrm{CX}$-based circuit construction predicts an asymptotic depth ratio of $\frac{4}{3}$ for single-layer QAOA~\cite{twine} and a ratio of 1 for two QAOA layers. However, compilation down to the pulse level actually reverses this trend, providing up to 25\% (35\%) depth reduction for $p=1$ ($p=2$) for Parity Twine QAOA over the reference implementation.

The resource estimates provided here serve as a connection between abstract algorithmic design and physical hardware constraints. In the development of EO-qubit chips, our results provide the necessary inputs to assess the feasibility of algorithm implementations for given a device. By integrating our pulse-level estimates with device-specific parameters, such as $T_2$ dephasing times, pulse duration, gate error rates, and the capacity for parallel pulse execution, our results allow to quantitatively predict the maximum system size achievable before measurement signals are lost. 

\section{Error Detection in Parity Twine\label{sec_error_detection}}

\begin{figure*}[t]
    \centering
    \includegraphics[width=.99\textwidth]{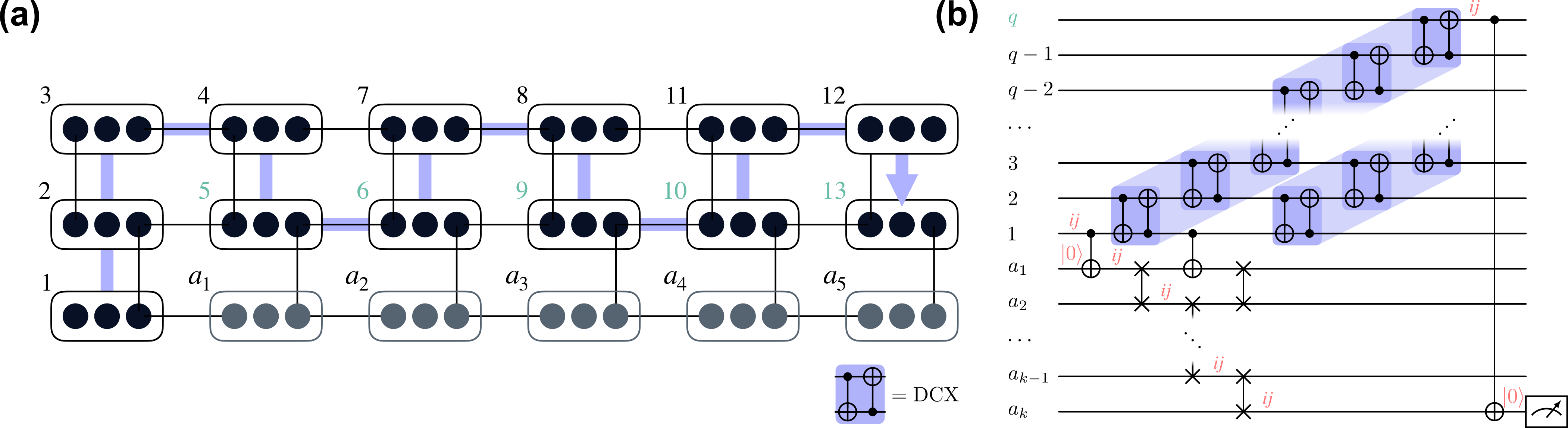}
    \caption{Illustration of the error detection scheme. (a) Topology of a tri-linear array of 18 EO qubits with the same connectivity and qubit placement as in the previous section. A Parity Twine-based algorithm is executed on 13 qubits using linear Twine along the purple path from qubit 1 to 13. Gray qubits $a_k$ are auxiliary qubits which are used to detect errors in the Twine chains which end in qubits $\mathcal D = \{ 5, 6, 9, 10, 13\}$ (teal). (b) Circuit diagram illustrating the protocol and relevant parity labels. Immediately before the start of a Twine chain ending in $q \in \mathcal D$ the gate $\mathrm{CX}_{1,a_1}$ duplicates the parity $Z_{ij}$ from qubit 1. The Twine chain (purple) commences with $\mathrm{DCX}$ gates while the duplicate label is transported to $a_k$ adjacent to $q$ by a chain of $\mathrm{SWAP}$ gates. A closing $\mathrm{CX}_{q, a_k}$ uncomputes the label. The shot is accepted or discarded conditional on the measurement of $a_k$. This procedure is repeated for every suitable Twine chain.} 
    \label{fig_ed_sketch}
\end{figure*}

In the previous section we studied the resource scaling of cornerstone quantum algorithms with physical EO qubits. However, the increasing number of imperfect gates leads to an accumulation of errors which effectively limits the size of a circuit. QEC promises to exponentially suppress errors as information is encoded in a growing number of qubits~\cite{Preskill1998,NielsenChuang,Preskill2018quantumcomputingin}, but as we will discuss in Sec.~\ref{sec_fermion_appl}, sufficiently large QEC codes may be unfeasible in the immediate future. In this section we explore an error detection technique native to Parity Twine as an intermediate step, introducing a moderate overhead to remove certain errors by postselection. Fig.~\ref{fig_ed_sketch} illustrates the protocol for a $3\times 6$ EO qubit array with the same assumptions as in the previous section .

In Parity Twine, a $\mathrm{DCX}$ chain starting in qubit 1 and ending in qubit $q$ transports the parity $Z_{ij}$ (or $Z_1$ at the start of the algorithm) from qubit 1 to $q$, as illustrated in Fig.~\ref{fig:QFT_circuits}(a)~\cite{twine,flow}. Connecting the start of a chain to its end on a secondary route provides the opportunity to detect bit-flip errors during its $\mathrm{DCX}$ gates: Immediately before the first $\mathrm{DCX}$, a gate $\mathrm{CX}_{1,a_1}$ duplicates the parity label to an auxiliary qubit $a_1$ which was initialized in the state $|0\rangle$. While the Twine chain proceeds, successive $\mathrm{SWAP}$ gates move the state of $a_1$ through an array of unused qubits to $a_k$ which is connected to $q$. If no bit-errors occurred during both the Twine and $\mathrm{SWAP}$ chains a gate $\mathrm{CX}_{q,a_k}$ after the final $\mathrm{DCX}$ returns the auxiliary qubit back to $|0\rangle$. This can be verified by measuring $a_k$. The experiment is discarded and repeated in a new shot if the measurement finds $a_k$ in state $|1\rangle$.

This protocol is readily implemented on a $3\times L$ square lattice as shown in Fig.~\ref{fig_ed_sketch}(a). The linear-chain Twine is used to implement an algorithm on a chain of $2L+1$ qubits, while the remaining $L-1$ qubits are dedicated to the role of auxiliary qubits $a_k$. The $3\times 6$ example allows error detection in the five Twine chains that end in qubits $q\in\mathcal D = \{5,6,9,10,13\}$. The circuit diagram is outlined in Fig~\ref{fig_ed_sketch}(b). For each of these chains, two $\mathrm{CX}$ gates are added to the circuit and the delay before the next Twine chain can start is increased by 1 $\mathrm{CX}$. The gate count of the PTN is thus increased by $2(L-1)$ $\mathrm{CX}$s and its depth by $L_1$ $\mathrm{CX}$s. Additionally, each auxiliary qubit is subject to $L-2$ $\mathrm{SWAP}$ gates. The measurements may increase the duration of a shot but no quantum operation depends on the result.

A peculiarity of the EO platform are leakage errors. The state of an EO qubit is defined by the spin quantum number of the three spins ($a, b, c)$. We follow Refs.~\cite{hrl_qpu, Russ2017} and define $|0\rangle$ as the singlet state with $S^{(a,b)} = 0$ and total spin $S = 1/2$ and $|1\rangle$ as a superposition of triplet states with $S^{(a,b)} = 1$, $S = 1/2$. The magnetic quantum number $m = \pm 1/2$ is inconsequential. These spin doublets span the computational space. The leakage space is spanned by the remaining triplet states of $a$ and $b$ combined with the two orientations of the gauge spin $c$, characterized by $S^{(a,b)} = 1$, $S = 3/2$.

Pauli spin blockade on spins $a$ and $b$ distinguishes singlet from triplet states, which provides one bit of classical information and is enough to measure a qubit, assuming no leakage has occurred. If no other information is gathered, a leaked state is indistinguishable from $|1\rangle$. We furthermore note that the encoded gates are in general not leakage-controlled, propagating leakage errors from target to control qubit, as further leakage or bit errors~\cite{Chadwick2025,hrl_qpu}. Consequently, a leakage error in the Twine or $\mathrm{SWAP}$ chain will result in a measurement indicating $|1\rangle$ and the shot being discarded.

We estimate the feasibility of the error detection protocol with a simple error model where each exchange pulse has the same error channel including both Pauli and leakage errors characterized by a single probability $p$. Each exchange pulse represents a Heisenberg interaction between the involved qubits, resulting in fully unbiased qubit errors where Pauli $x$-, $y$-, and $z$-channels are applied with equal probability $p/4$, the remaining $p/4$ is assigned to leakage errors~\cite{hrl_qpu}. Our error scheme can detect Pauli-$x$ and -$y$ errors, which result in a bit-flip, and leakage errors, as argued above, while $z$ errors are undetectable.

We further assume that errors on different exchange pulses are independent, and model the errors along a sequence of $n_p$ pulses with a binomial distribution. The probability
\begin{align}
    P_{f,\mathcal{S}} (n_p) = \sum_{e\in \mathcal{S}} \binom{n_p}{e} (f p/4)^e (1-fp/4)^{n_p-e}
\end{align}
describes the occurrence of a set $\mathcal{S}$ of error counts. Relevant cases are $\mathcal{S} = \{0\}$ for no errors, $\mathcal{S} = \mathrm{even} (\mathrm{odd}$) for an even (odd) number of errors, and $\mathcal{S} = \mathrm{even}>0$ for a non-zero even number. The integer $f$ distinguishes between detectable errors ($f=3$), undetectable errors ($f=1$), and arbitrary error ($f=4$), respectively. The pulse count $n_p(q)$ corresponds to the pulse sequence length of $q-1$ $\mathrm{DCX}$ gates of a Twine chain and $n_p(a)$ corresponding to four $\mathrm{SWAP}$ and two $\mathrm{CX}$ gates for the auxiliary chain. For simplicity, we consider a clean Parity Twine network without single qubit gates between the chains.

\begin{figure}[t]
    \centering
    \includegraphics[width=.99\columnwidth]{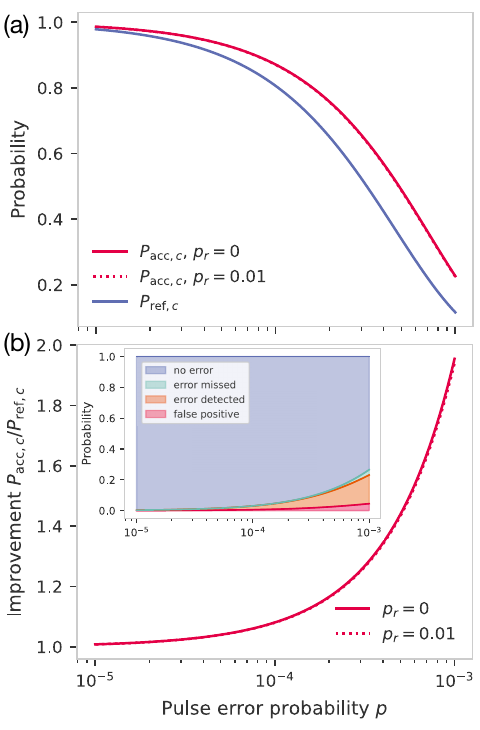}
    \caption{Performance of Parity Twine with error detection on the chip from Fig.~\ref{fig_ed_sketch}(a) as a function of the pulse error rate $p$. (a) Probability for the accepted result being correct, $P_{\mathrm{acc},c}$ (red) without (solid) and with (dotted) measurement error $p_r$. The purple line is the reference of a single shot without error detection.
    (b) Ratio of $P_{\mathrm{acc},c} / P_{\mathrm{ref},c}$. In the depicted regime the improvement grows rapidly with increasing error rate. With further increasing $p$ the ratio will eventually decay again due to undetectable errors. 
    Inset: Distribution of probabilities for the different outcomes of the Twine chain ending in $q = 13$, setting $p_r=0$. The largest share of the probability corresponds to the correct outcome throughout, detected error events become more prominent with increasing $p$. Missed errors and false positives are generally low. A shot is accepted in the purple and teal area.}
    \label{fig_ed_performance}
\end{figure}

The error detection in a single Twine chain ending in qubit $q\in \mathcal D$ can have the following four outcomes. (i) No detectable error occurs in the Twine or auxiliary chain, correctly signaling acceptance,
\begin{align}
    P_\mathrm{corr} (q) = P_{3,\{0\}}(n_p(q)) P_{3,\mathrm{even}}(n_p(a)).
\end{align}
(ii) A detectable error in the Twine chain may be missed because it is masked by an error in the auxiliary chain or a subsequent error restoring the bit value in a later pulse, wrongly signaling acceptance,
\begin{align}
    P_\mathrm{miss} (q) =& P_{3,\mathrm{odd}}(n_p(q)) P_{3,\mathrm{odd}}(n_p(a))\nonumber\\
    &+ P_{3,\mathrm{even}>0}(n_p(q)) P_{3,\mathrm{even}}(n_p(a)).
\end{align}
(iii) An odd number of errors in the Twine chain is detected, and some masked errors may accidentally be discovered due to an independent error in the auxiliary chain, the shot is correctly rejected,
\begin{align}
    P_\mathrm{dect} (q) =& P_{3,\mathrm{odd}}(n_p(q)) P_{3,\mathrm{even}}(n_p(a))\nonumber\\
    &+ P_{3,\mathrm{even}>0}(n_p(q)) P_{3,\mathrm{odd}}(n_p(a)).
\end{align}
(iv) An error in the auxiliary chain may also lead to a false positive, an otherwise correct shot is wrongly rejected,
\begin{align}
    P_\mathrm{fals} (q) = P_{3,\{0\}}(n_p(q)) P_{3,\mathrm{odd}}(n_p(a)).
\end{align}
These probabilities are illustrated for the longest Twine chain with $q=13$ in the inset of Fig.~\ref{fig_ed_performance}(b), demonstrating a high acceptance rate and a low chance of missing errors. 

We model measurement errors as bit-flip channel with a probability $p_r$ during the readout. This channel can obscure the events described above and lead to a higher chance of false positives,
\begin{align}
    P'_\mathrm{corr} (q) =& (1-p_r) P_\mathrm{corr} (q), \\
    P'_\mathrm{miss} (q) =& (1 -p_r) P_\mathrm{miss} (q) + p_r P_\mathrm{dect} (q),\\
    P'_\mathrm{dect} (q) =& (1-p_r) P_\mathrm{dect} + p_r P_\mathrm{miss} (q), \\
    P'_\mathrm{fals} (q) =& (1-p_r) P_\mathrm{fals} (q) + p_r P_\mathrm{corr} (q) .
\end{align}
Independent of the detectable errors, phase errors in the chains with detection are undetectable, as well as any errors in the other Twine chains,
\begin{align}
    P_\mathrm{unde} (q) = P_{f, \leq n_q}(n_p(q))
\end{align}
with $f = 1$ if $q\in\mathcal D$ and $f = 4$ otherwise.

A shot of the full PTN is accepted, if all measurements indicate acceptance, either due to a correct outcome or a masked error. The total acceptance probability is
\begin{align}
    P_\mathrm{acc} = \prod_{q\in\mathcal D} \left(P'_\mathrm{corr}(q) + P'_\mathrm{miss}(q)\right).
\end{align}
We further define the conditional probability, that an accepted result is correct, taking both missed detectable and undetectable errors into account,
\begin{align}
    P_{\mathrm{acc},c}  = \left(\prod_q\left(1 - P_\mathrm{unde}(q)\right)\right)\left( \prod_{q\in\mathcal D} P'_\mathrm{corr}(q) \right) / P_\mathrm{acc}.
\end{align}
Here, $P_{\mathrm{acc},c}$ describes the probability that the correct quantum state was prepared and accepted by the algorithm. As a reference, we compare to the probability of obtaining the correct result in a single shot with no error detection, $P_{\mathrm{ref},c} = \prod_q P_{4, \{0\}} (q)$.

The probabilities $P_{\mathrm{acc},c}$ and $P_{\mathrm{ref},c}$ for the 18 EO chip are plotted in in Fig.~\ref{fig_ed_performance}(a). No improvement $P_{\mathrm{acc},c} / P_{\mathrm{ref},c}$ is obtained for low error rates but it becomes substantial in the experimentally relevant regime~\cite{hrl_qpu}. The improvement peaks and then rapidly decays since both $P_{\mathrm{acc},c}$ and $P_{\mathrm{ref},c}$ eventually fall to 0 due to the undetectable errors. The position of the maximum depends on the number of pulses, here it is found at $p\approx 4.98 \times 10^{-3}$, which is far outside the experimentally relevant regime and thus not shown.

\begin{figure}[t]
    \centering
    \includegraphics[width=.99\columnwidth]{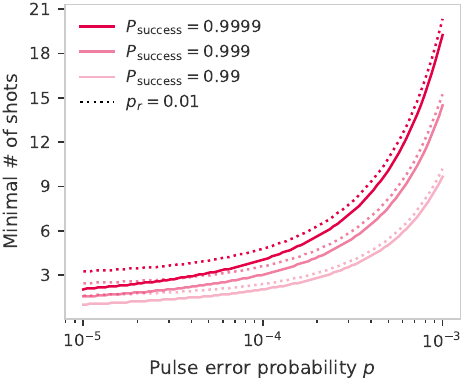}
    \caption{Minimal number of shots as a function of the pulse error rate $p$ on the chip from Fig.~\ref{fig_ed_sketch}(a). The curves indicate the number of shots after which a fraction given by $P_\mathrm{success}$ of experiments can be expected to be concluded successfully. For high error rates, a diverging number of shots is required to guarantee the improvement suggested by Fig.~\ref{fig_ed_performance}. In the intermediate regime, we find a feasible overhead complementing the good improvement of $P_{\mathrm{acc},c}$. Dotted lines include a small measurement error, which becomes relevant at low pulse error rates.} 
    \label{fig_ed_num_shots}
\end{figure}

The price for the improvement is an overhead in shots. To quantify this overhead, we define the success probability as the probability that at least one of $n_s$ shots was accepted,
\begin{align}
    P_\mathrm{success}(n_s) = 1 - (1 - P_\mathrm{acc})^{n_s}.
\end{align}
We then invert this relation to estimate the minimum number of shots required for an experiment to conclude successfully. The result is plotted in Fig.~\ref{fig_ed_num_shots}.

While small measurement errors have a limited impact on $P_{\mathrm{acc},c}$, the increased rate of false positives becomes relevant at low pulse error rates, where only a low improvement is expected. At high pulse error rates, the required number of shots diverges to prohibitive numbers. In the intermediate regime, where we predicted a substantial improvement, however, we find a feasible overhead. In particular, we find an improvement of $P_{\mathrm{acc},c}/ P_{\mathrm{ref},c} \approx (1.21, \ 1.16,\ 1.12)$ requiring (5-6, 4-5, 3-4) shots at a pulse error rate of $(2.5,\ 2,\ 1.5)\times 10^{-4}$.

We emphasize that this promising outcome is enabled by the strengths of the EO platform and can not be expected from other qubit modalities. Specifically, the low pulse count of a $\mathrm{SWAP}$ gate entails few errors in the auxiliary chain, a major reason for the low probability of false positives and of missing an error, making the accepted shots likely correct. The unique error model including leakage errors as a detectable source of infidelity is another asset, leaving only a small fraction of error events undetectable. 

Our model follows experimental data suggesting that leakage errors are three times less likely than qubit errors. If leakage becomes more prevalent, we expect the results to improve further. On the other hand, we also assumed that all leakage errors spread along the Twine chain. If the gates become more leakage-controlled and contain the leakage, we expect the improvement to be reduced. As long as leakage errors cannot be suppressed altogether, the non-leakage-controlled nature of the gates is in fact helpful to make sure the errors arrive at the auxiliary qubit.

\begin{figure}[t]
    \centering
    \includegraphics[width=.99\columnwidth]{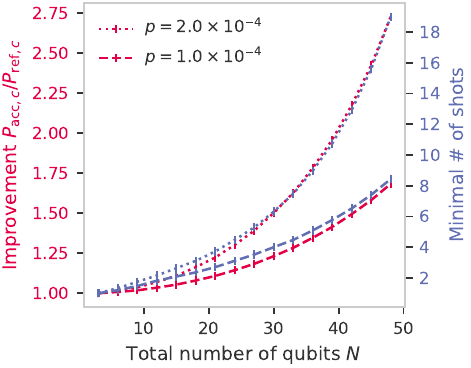}
    \caption{Scaling of the error detection method as a function of the qubit number $N$. We plot the improvement $P_{\mathrm{acc},c} / P_{\mathrm{ref},c}$ (red) and the expected number of shots (purple) for two different values of the pulse error rate $p$ and $P_\mathrm{success} = 0.999$, assuming $p_r = 0.01$. As $N$ grows, the number of shots increases exponentially, making the method not arbitrarily scalable despite the promising results for low qubit numbers. The growth is slower for lower error per pulse $p$, determining how many qubits can be tackled.
    Note that $N$ includes auxiliary qubits, only $2N/3 + 1$ qubits are used for the Parity Twine network.}
    \label{fig_ed_scaling}
\end{figure}

The encouraging results above were computed for the chip from Fig.~\ref{fig_ed_sketch}(a)  with 18 EO qubits. In Fig.~\ref{fig_ed_scaling} we investigate the scaling of the method. We consider a $N = 3 \times L$ array of EO qubits following the same layout pattern with $\mathcal D = \{q > 2\ \text{with}\ (q-1)\mod 4 \leq 1\}$. The improvement $P_{\mathrm{acc},c} / P_{\mathrm{ref},c}$ grows with the number of qubits, indicating that $P_{\mathrm{acc},c}$ decays slower than $P_{\mathrm{ref},c}$. This is the expected behavior since certain errors are removed after error detection. However, at the same time, the number of required shots grows exponentially with $N$ or an exponentially lower error rate $p$ is required. This is expected due to the exponential decay of the probability $p_\mathrm{corr}$. A geometry that allows for a larger set $\mathcal{D}$ to reduce undetectable errors, e.g., a $2\times L$ ladder, will increase the improvement but does not prevent the decay of $p_\mathrm{corr}$. Error detection is thus not scalable to arbitrary qubits numbers, however, it promises good results for low to intermediate qubit numbers. Depending on $p$ a major improvement can be expected for up to tens of qubits with a moderate overhead.

Finally, we note that we considered a clean Parity Twine network which generates all two-qubit parity labels and overwrites them without applying any single-qubit gates. Thus, we have studied the scaffolding of an arbitrary quantum algorithm requiring all pairwise interactions, for example QFT or QAOA for dense problems. Single qubit gates between the Twine chains mostly contribute to the undetectable errors, but add only a limited number of pulses, on the 18 EO chip roughly equivalent to one additional $\mathrm{CX}$ gate per two Twine chains. Therefore, we do not expect a major impact on the results and conclude that these results represent an approximate estimate for an arbitrary quantum algorithm realized by Parity Twine.

\section{Towards Fault-Tolerance: Simulation of Fermion Lattices\label{sec_fermion_appl}}

In this section we turn to another application which promises to bridge proof-of-principle demonstration and the NISQ-regime to practical application and the era of fault-tolerant quantum computing: The simulation of fermion systems, which is of great relevance for a plethora of applications in research and development~\cite{doi:10.1073/pnas.1619152114,PhysRevLett.120.110501,PRXQuantum.5.030323,Alam2025,PhysRevA.79.032316,Maskara2025,PhysRevX.8.011044,SqDRIFT,aigner_fermion_2026}. The interactions on fermion lattices are the microscopic origin of many complex macroscopic phenomena in solid-state physics and fermionic interactions are a vital ingredient to simulating the electronic structure of molecules in quantum chemistry. Spin-$1/2$ qubits encoded in the spins of single particles provide a natural platform for analogue simulations~\cite{Dehollain2020,PhysRevResearch.4.033043,Kiczynski2022,Morozova2026}. However, this procedure is incompatible with QEC, and thus limited to small-scale applications and quantitative observations.

Digital simulations of fermionic systems typically require a fermion-to-qubit mapping~\cite{BRAVYI2002210,Verstraete_2005,PhysRevB.104.035118}, of which the Jordan--Wigner (JW) encoding is the most prominent example. It represents fermionic operators as qubit operators which are placed in a rigid, one-dimensional order. Consequently, simulating higher-dimensional lattices incurs a significant overhead due to non-local interactions, which are commonly mitigated by dynamically reordering fermionic modes through fermionic SWAP networks (FSNs). In a straightforward application, FSNs induce an $\mathcal O (\sqrt N)$ overhead in depth and gate count for simulations of two-dimensional lattices~\cite{PhysRevLett.120.110501, PRXQuantum.5.030323,Alam2025}.

The overhead of the FSN can be reduced on reconfigurable qubit arrays at the cost of introducing ancillas~\cite{Maskara2025} or by leveraging fully connected qubit systems~\cite{Constantinides2025}. Such techniques may be enabled by coherent spin shuttling~\cite{Struck2024,DeSmet2025,Unseth_2026_parity}. However, shuttling has not yet been explored for EO qubits and the demand for auxiliary qubits or all-to-all shuttling of logical qubits can be challenging~\cite{Siegel_2026_snakes,Eggerickx_2026}.

Therefore, we consider a recently discovered method~\cite{aigner_fermion_2026} which is derived from a generalization of Parity Twine~\cite{flow}. It overcomes the one-dimensional connectivity restriction by switching to a complementary second JW encoding. Effectively, this enables fermionic modes that would be distant under a single JW encoding to interact using a second physical dimension of a qubit lattice. This lowers the resource cost for simulating higher-dimensional fermion lattices and for routing of fermions.

In this section, we investigate two applications from the field of fermion simulation: the fermionic fast Fourier transform (FFFT) in Sec.~\ref{sec_ffft}, an algorithm with low requirements which can potentially be useful on a similar scale as the QFT, and the simulation of a Fermi-Hubbard model in Sec.~\ref{sec_lattice_surgery}, which likely requires QEC. In both cases we study the parity method and the respective state-of-the-art. The results are promising for benchmarks on a small scale and indicate which improvements are needed for QEC.

\subsection{FFFT as Near-Term Algorithmic Benchmark\label{sec_ffft}}

First, we discuss the efficient implementation of the fermionic fast Fourier transform (FFFT) with EO qubits. It is a powerful tool in digital quantum simulation~\cite{PhysRevA.79.032316}. While in momentum space, translationally invariant free-fermion states can be prepared in constant depth. The FFFT subsequently transforms them to real space, making it relevant for preparing states with long-range structure, including chiral superconductors, critical states, and Fermi surfaces~\cite{Maskara2025}. Furthermore, the FFFT can enable Hamiltonian simulation of periodic materials~\cite{PhysRevX.8.011044}.

\begin{figure}[t]
    \centering
    \includegraphics[width=.99\columnwidth]{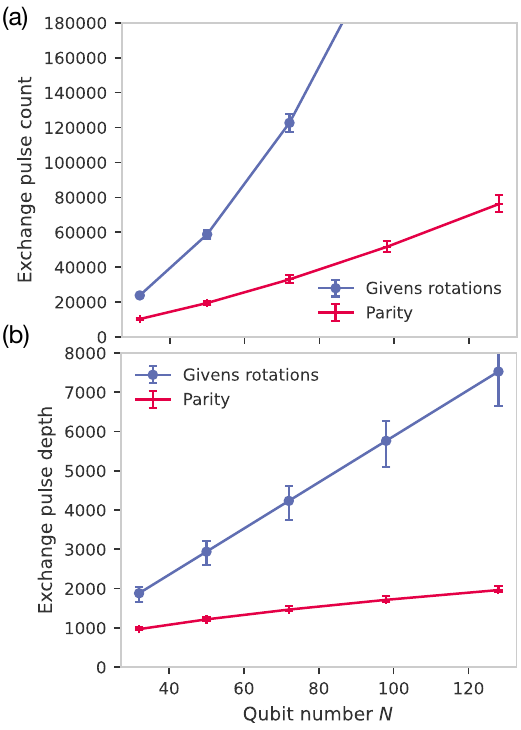}
    \caption{Resource estimates for a two-dimensional FFFT on a $N = L\times 2L$ square lattice of qubits. Not only has ParityQC's approach (red)~\cite{aigner_fermion_2026} a better asymptotic scaling in both exchange pulse count and depth than the Givens rotations (purple)~\cite{PhysRevLett.120.110501}, it is favourable throughout, starting from the recently reported 18 qubits~\cite{hrl_qpu}. The error bars indicate the range of each data point due to the arbitrary configurations of the EO qubits, the depth values are obtained with maximal parallelization during each gate.}
    \label{fig_fermion_ffft}
\end{figure}

We estimate the resource requirements for a two-dimensional FFFT considering both the recent parity strategy based on dynamic encoding~\cite{aigner_fermion_2026} and a realization via the Givens rotations as previous state of the art~\cite{PhysRevLett.120.110501} on EO qubits. We translate circuit implementations from Ref.~\cite{aigner_fermion_2026} and Ref.~\cite{PhysRevLett.120.110501} into exchange pulse sequences based on Ref.~\cite{Chadwick2025}. As illustrated in Fig.~\ref{fig_fermion_ffft}, for small systems the total resources are comparable to the Twine QFT discussed in Sec.~\ref{near_term}. With the parity method the FFFT on a $3\times 6$ square grid requires $\approx 4500$ exchange pulses in a depth of $\approx 700$ pulses, assuming maximal parallelization, which is 38\% and 32\% less compared to the Givens rotations, respectively. The variation of sequence lengths of individual gates due to the internal configuration of spins in each EO qubit is indicated by the error bars in Fig.~\ref{fig_fermion_ffft}. It has a minor effect, although in the case of deep circuits a dedicated compilation pass optimizing the initial configurations before the algorithm may be valuable.

The parity approach achieves a remarkable asymptotic resource scaling. It requires $\mathcal O (N^{3/2})$ pulses in a depth of $\mathcal O (\sqrt{N})$ compared to $\mathcal O (N^2)$ pulses in a depth of $\mathcal O (N)$ for the Givens rotations. The FFFT thus scales favorable compared to the QFT and is a promising candidate for early proof-of-principle algorithmic benchmarks with physical qubits, if no error detection as described in Sec.~\ref{sec_error_detection} is required.

\subsection{Partially Fault-Tolerant Trotter Simulation of a Fermi-Hubbard Model\label{sec_lattice_surgery}}

\begin{figure}[t]
    \centering
    \includegraphics[width=.99\columnwidth]{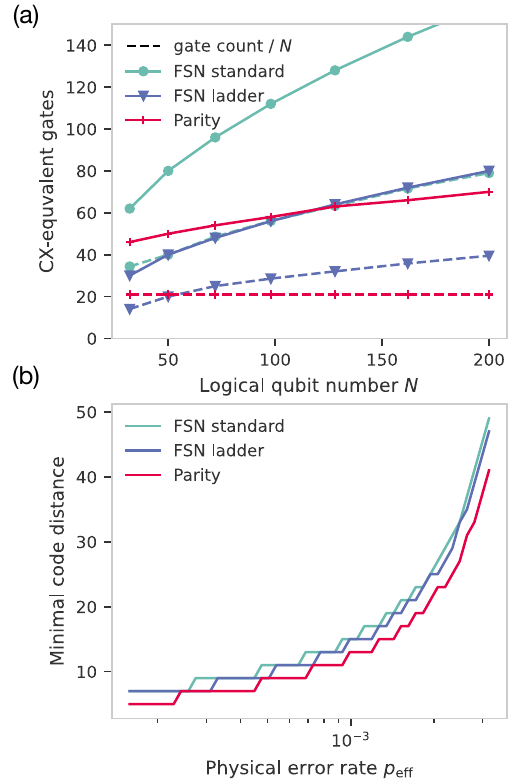}
    \caption{Requirements for simulating a single second-order Trotter step of a 2D Fermi-Hubbard model with nearest-neighbor interactions on a square lattice of $N = L\times 2L$ (logical) qubits.
    (a) Two-qubit gate depth (solid) and count (normalized by $N$ for clarity, dashed) for ParityQC's approach~\cite{aigner_fermion_2026}, a standard FSN~\cite{PhysRevLett.120.110501} and an optimized version on a ladder connectivity graph~\cite{PRXQuantum.5.030323,Alam2025}. When using lattice surgery, the depth of ParityQC's approach becomes constant.
    (b) Minimal code distance $d$ required to realize the required depth of logical gates in one Trotter step with fidelity $F_\mathrm{target} = 0.99$ and $N = 128$ logical qubits, using the algorithms from panel (a) as a function of the effective physical error rate $p_\mathrm{eff}$. The code distance is very sensitive to $p_\mathrm{eff}$ and diverges near the threshold $p_\mathrm{th} = 5.7\times 10^{-3}$, such that much lower error rates will be required.} 
    \label{fig_fermion_lattice}
\end{figure}

Despite the advantageous scaling, utility scale applications challenge the boundaries of what can be achieved with physical qubits. At least 108 (logical) qubits are necessary for simulating the electronic structure of a molecule related to fertilizer production~\cite{doi:10.1073/pnas.1619152114}, at which point unfeasible error rates per pulse are demanded. Fig.~\ref{fig_fermion_lattice}(a) illustrates the circuit depth and gate count of a single second-order Trotter step simulating a two-dimensional Fermi-Hubbard model with nearest-neighbor-interaction using the parity method or an FSN. A few Trotter steps with a few qubits may serve as a near-term benchmark comparable to the FFFT, however, the numbers required for practical utility can be anticipated to be well beyond the capabilities of current physical qubits.

QEC and fault-tolerant quantum computing promise to suppress the errors of encoded logical qubits far below the physical error rate, enabling significantly larger computations. In the following, we aim to predict requirements for a simulation of the Fermi-Hubbard model in the framework of QEC and to identify critical limitations that disproportionally contribute to the overhead. We first estimate the minimal code distance for each algorithm, and then relate it to the performance of EO qubits to assess the feasibility of a realization depending on the local qubit parameters.

We assume a surface code on a square grid of EO qubits~\cite{Dennis2002_topological,PhysRevA.86.032324} that alternate between data and measurement qubits, where the latter are used for stabilizer measurements and leakage removal protocols (LRPs)~\cite{hrl_qpu}. This choice is motivated by the fact that the dynamic switching between complimentary JW encodings in the parity approach can be performed via lattice surgery in $\mathcal O (1)$ depth~\cite{aigner_fermion_2026}.

Lattice surgery is a way to perform fault-tolerant operations in a surface code~\cite{Campbell_2017_fault_tolerant,Litinski2019gameofsurfacecodes}. It relies on the fact that measuring the stabilizers between boundaries of logical qubits merges these boundaries and induces a joint Pauli measurement of the logical qubits. Reversely, a surface code can be split into two by measuring a line of data qubits. A logical $\mathrm{CX}$ can be implemented by merging and splitting the control with an auxiliary logical qubit prepared in $|+\rangle$ in a $ZZ$ measurement, then merging and splitting the auxiliary qubit with the target in a $XX$ measurement. A subsequent $Z$ measurement of the auxiliary qubit removes it and may impose $Z$ or $X$ corrections on control and target.

Using lattice surgery, the encoding switch has a depth of three logical measurements and four layers of single-qubit Clifford gates~\cite{aigner_fermion_2026}. The fermionic interaction terms are constant in depth, too, and add four additional layers of logical $\mathrm{CX}$ gates and a limited number of single-qubit gates to each Trotter step. Single-qubit Clifford gates can be implemented virtually by classical tracking~\cite{Litinski2018latticesurgery}, therefore we only count two layers of non-Clifford gates, one in each JW-encoding. For FSNs we count the $\mathcal{O}(\sqrt N)$ depth from Fig.~\ref{fig_fermion_lattice}(a) and add the interaction terms.

In fault-tolerant quantum computing, non-Clifford gates represent the main cost of an algorithm. Single-qubit non-Clifford gates can be approximated to arbitrary precision with $T$ gates at a cost determined by the Solovay--Kitaev theorem, requiring up to 120 $T$ gates for practical error rates~\cite{NielsenChuang}. A $T$ gate is constructed by teleporting a magic state into the logical qubit~\cite{PhysRevA.62.052316}, while the magic states are obtained from a distillation protocol~\cite{Litinski2019magicstate,Gidney2019efficientmagicstate}. The cost of a single magic state can reach up to tens of thousands of physical qubits and tens of stabilizer measurement cycles~\cite{Litinski2019magicstate}. These numbers can be reduced if a decreased fidelity of the magic states is acceptable, for example by using magic state cultivation, but the dominant role of the $T$ gate cost remains unchallenged~\cite{Thorvaldson2026}. We consider this to be elusive for the near future.

To facilitate an earlier application, we instead assume a partially fault-tolerant protocol where analogue rotations are injected into the logical state in two steps of lattice surgery on average~\cite{PRXQuantum.5.010337}. The analogue rotations rely on resource states which are teleported into the logical qubits. Errors in the preparation of these states are not correctable. To avoid a limiting effect on the total fidelity, the accumulated error of $\mathcal O(N)$ rotations should be small compared to the error of the lattice surgery operations. Thus, either some additional overhead for preparing high-fidelity resource states or a compromise on the total fidelity can be anticipated. In the case of low-fidelity resource states, certain expectation values can still be sampled with high accuracy~\cite{PRXQuantum.5.010337} and the code distance can be truncated to match the limited logical fidelity. In the following, we assume a sufficiently high fidelity of resource states.

The circuit depth determines the number of successive stabilizer measurement rounds $\nu$. Per round, the logical error is $p_{L,\mathrm{round}} \approx C \left( p_\mathrm{eff} / p_\mathrm{th}\right)^{(d+1)/2}$, given the effective physical error rate $p_\mathrm{eff}$ per operation, the threshold error rate $p_\mathrm{th}$, code distance $d$, and the geometry-dependent coefficient $C \approx 0.03$~\cite{PhysRevA.86.032324}. With an error model appropriate for EO qubits, depolarizing noise on data and measurement qubits, and initialization, gate, and measurement errors under MWPM decoding, $p_\mathrm{th} \approx 5.7\times 10^{-3}$ was found~\cite{PhysRevA.86.032324}. To mitigate the effect of measurement errors all stabilizer measurement must be repeated several times~\cite{Dennis2002_topological}. Since the measurement fidelity is limited in spin qubits, we assume $d$ repetitions, equal to the code distance. Thus, the cumulative probability of a logical failure per lattice surgery operation per qubit is $p_{L,\mathrm{surgery}} \approx 1 - (1 - p_{L,\mathrm{round}})^d \approx  d p_{L,\mathrm{round}}$.

To complete one Trotter step with $N$ logical qubits with a target fidelity $F_\mathrm{target}$, we demand
\begin{align}
    (1 - p_{L,\mathrm{surgery}})^{N\nu} \geq F_\mathrm{target}
\end{align}
where $\nu$ is the stabilizer measurement depth. We solve for the minimal required code distance as a function of physical error rates $p_\mathrm{eff}$, the result is shown in Fig.~\ref{fig_fermion_lattice}(b) for $F_\mathrm{target} = 0.99$, $N = 128$. The number of physical qubits per logical qubit is then determined by the code distance as $N_\mathrm{phys} / N = 2d(d - 1) + 1$
. Thanks to its improved scaling, the parity approach loosens the requirements for QEC, in particular for high $p_\mathrm{eff}$ where $d$ is highly sensitive. For example, for $p_\mathrm{eff} = 2\times 10^{-3}$ the parity approach results in 841 physical qubits needed to encode one logical qubit, 30\% (40\%) fewer than the ladder (standard) FSN.

Depending on $p_\mathrm{eff}$, a few hundreds to thousands of physical qubits per logical qubit are required. Naturally, it is critical to relate the effective parameter $p_\mathrm{eff}$ to a more hardware-realistic error model. We model the effective error as a linear combination
\begin{align}
    p_\mathrm{eff} = \alpha p_g + \beta p_l + \gamma p_m + \delta p_i
\end{align}
of errors stemming from gates $p_g$, the leakage removal protocol $p_l$, measurement $p_m$, and idling times $p_i$~\cite{Google2021,Google2023,Google2025}. During a stabilizer measurement cycle each data qubit is involved in four physical $\mathrm{CX}$ gates with different measurement qubits and idles during the measurement. Additionally, it participates in a LRP, which is a unique requirement of EO qubits.

We count each gate, idling period, LRP and measurement as a circuit location and weigh each error with the share of locations it is associated with. Next, we decompose gates and LRP into pulse sequences of length $n_p(\mathrm{CX})$ and $n_p(\mathrm{LRP})$, respectively. The pulse error probability $p$ then accumulates during a pulse sequence as $p_{g(l)} = 1 - (1 - p)^{n_p} \approx n_p p$. Further modelling $p_i \approx (T_\mathrm{meas} / T_2)^2$ with the measurement duration $T_\mathrm{meas}$ and coherence time $T_2$, we thus obtain
\begin{align}
    p_\mathrm{eff} \approx \left(\frac{4}{7} n_p(\mathrm{CX}) + \frac{1}{7} n_p(\mathrm{LRP})\right) p + \frac{p_m}{7} + \frac{1}{7}\left(\frac{T_\mathrm{meas}}{T_2}\right)^2.
\end{align}

We use  parameters $p\approx 1.5\times 10^{-4}$, $p_m \approx 0.01$, and $T_2 \approx \SI{19.3}{\micro\second}$ and the qubit connectivity map reported in Ref.~\cite{hrl_qpu}, and furthermore assume $n_p(\mathrm{CX}) = 24.3$ and $n_p(\mathrm{lrp}) = 24$ on average. With the optimistic assumption of $T_\mathrm{meas} = \SI{300}{\nano\second}$ we find $p_\mathrm{eff} \approx 4.06 \times 10^{-3}$ as a result. Of this error, $\approx 64\%$ relates to the pulse error $p$ due to the large number of relevant circuit locations, while the large measurement error is responsible for $\approx 35\%$, and idling during the measurement for $\approx 1\%$. While the obtained value for $p_\mathrm{eff}$ is below the threshold, based on our model we a expect a minimal code distance far beyond realistic hardware possibilities in the foreseeable future.

The dominant role of $p$ motivates architectural optimization, since the connectivity between individual quantum dots has a strong impact on the pulse sequence length of encoded gates. EO qubits defined on a triangular lattice allow for significantly shorter gates~\cite{Chadwick2025,Broz2025}. With this layout we find $p_\mathrm{eff} \approx 3.5 \times 10^{-3}$ due to the shorter $n_p(\mathrm{CX})$, even if $n_p(\mathrm{LRP})$ is unchanged. The corresponding code distance of $d = 51$ is still unrealistic, however, this corresponds to a $51\%$ reduction in the number of physical qubits without improving operation fidelities, which highlights the importance of the chip layout.

We can thus conclude, that while error rates may be sufficiently low for small proof-of-principle QEC demonstrations, the required overhead in qubit number is still prohibitive for practical applications. Our results provide target error rates as guidance for further development, reducing the number of pulses or the error per pulse seems the most efficient path to improvements. Further simplifications on the path the fault-tolerance beyond the scope of our estimates can potentially be achieved by an informed choice of codes~\cite{PhysRevA.109.032433,PRXQuantum.5.040328}, the shuttling of logical qubits~\cite{Siegel_2026_snakes,Chadwick2025_QEC} or error detection on the level of the EO encoding~\cite{Heinz2026}. For certain types of spin qubits, an exploitation of a noise bias is possible, which may offer benefits compared to the EO qubits discussed here~\cite{PhysRevA.109.032433,replacement_type,Siegel_2026_erasure,Thorvaldson2026}. We furthermore confirm the parity approach as a promising candidate for early fault-tolerant quantum simulation. However, we note that when transitioning to a fully fault-tolerant implementation the $T$ gate cost, which is $\mathcal{O}(N)$ for all methods, becomes dominant over the encoding switch, levelling differences and steeply increasing the requirements.

\section{Summary and Conclusion\label{sec_summary}}

In this paper we explored increasingly demanding and complex applications for EO spin qubits. Specifically, we derived resource requirements for the NISQ algorithm QAOA, the cornerstone quantum algorithms QFT and FFFT which can serve as intermediate-scale algorithmic benchmarks, and the utility-scale simulation of fermion lattices. We furthermore proposed a tailored error detection technique to cover the gap between algorithms with purely physical qubits and QEC.

By compiling the Parity Twine network to the hardware level, we showed that Parity Twine offers substantial synergies with EO qubits. Most notable is the ability of the EO platform to perform $\mathrm{DCX}$ gates directly. We identified the optimal representation of the Twine network and further improved the circuits by pulse-level scheduling with an increasing degree of parallelization for the examples of QFT and QAOA. The Parity Twine circuits itself already outperform other state-of-the-art compilation methods and their circuit structure remarkably offers a higher potential for depth reductions on the pulse level. Our findings show that EO QPUs with the recently demonstrated capabilities~\cite{hrl_qpu} are a prime candidate for near- and mid-term realizations of quantum algorithm based on the Parity Twine method. 

Building on these results, we presented an error detection technique which is native to the Parity Twine network. Harnessing the unique properties of the EO qubits we showed that error detection and post-selection can significantly improve the quality of the final quantum state. Our analysis includes leakage errors which can also be detected with a high probability. While not arbitrarily scalable, the overhead of the error detection remains low to moderate in the currently relevant regime. With lower error rates on maturing hardware, error detection can grow with the qubit number an remain relevant until the leap to quantum error corrected devices is feasible. Thus, Parity Twine equipped with error detection is a promising candidate for NISQ applications on a larger scale than permitted by physical qubits alone.

Moving beyond Parity Twine, we studied the implementation of a recent technique for the efficient digital simulation of fermions on EO qubits. Our resource estimates suggest that both FFFT and the Hamiltonian simulation of fermion lattices can serve as demonstrations as qubit numbers grow and especially the FFFT benefits from a favorable asymptotic scaling.  However, we expect the requirements for practical applications in quantum chemistry with around 110 (logical) qubits to be unrealistic with physical qubits.

Turning to QEC, we discussed a partially fault-tolerant implementation of the Fermi-Hubbard model, abstaining from magic state distillation and injection in favor of an earlier feasibility. We estimate the minimal code distance for this algorithm as a function of an effective physical error rate which we relate to the error model of EO qubits. While the estimates suggest that current fidelities are unsuitable for large-scale applications, the results identify the bottlenecks and suggest improvement strategies.

Our results emphasize the importance of hardware-informed compilation, provide experimental target values for the development of future generations of QPUs, and highlight the potential of EO qubits. While further work is needed to identify an optimal architecture for partially and fully fault-tolerant quantum computing with EO qubits, many applications are immediately feasible to illustrate the capabilities of spin qubits now and in the future.

\begin{acknowledgments}
The authors acknowledge insightful discussions with Thaddeus Ladd. We further thank Christophe Goeller, Gregor Aigner, Berend Klaver, Nitica Sakharwade, and Michael Fellner for helpful discussion and valuable feedback on earlier stages of the work and on the manuscript.
\end{acknowledgments}


\bibliography{literature}

@misc{hrl_qpu,
  author = {{Members of the HRL Quantum Team}},
  title = {A digitally controlled silicon quantum processing unit},
  year={2026},
  eprint={2604.16216},
  archivePrefix={arXiv},
  primaryClass={quant-ph}
}

@article{madzik_operating_2025,
	title = {Operating two exchange-only qubits in parallel},
	volume = {647},
	issn = {1476-4687},
  journal = {Nature},
	url = {https://doi.org/10.1038/s41586-025-09767-5},
	doi = {10.1038/s41586-025-09767-5},
	pages = {870--875},
	number = {8091},
	journaltitle = {Nature},
	shortjournal = {Nature},
	author = {Mądzik, Mateusz T. and Luthi, Florian and Guerreschi, Gian Giacomo and Mohiyaddin, Fahd A. and Borjans, Felix and Chadwick, Jason D. and Curry, Matthew J. and Ziegler, Joshua and Atanasov, Sarah and Bavdaz, Peter L. and Connors, Elliot J. and Corrigan, J. and Ercan, H. Ekmel and Flory, Robert and George, Hubert C. and Harpt, Benjamin and Henry, Eric and Islam, Mohammad M. and Khammassi, Nader and Keith, Daniel and Lampert, Lester F. and Mladenov, Todor M. and Morris, Randy W. and Nethwewala, Aditi and Neyens, Samuel and Otten, René and Osuna Ibarra, Linda P. and Patra, Bishnu and Pillarisetty, Ravi and Premaratne, Shavindra and Ramsey, Mick and Risinger, Andrew and Rooney, John D. and Savytskyy, Rostyslav and Watson, Thomas F. and Zietz, Otto K. and Matsuura, Anne Y. and Pellerano, Stefano and Bishop, Nathaniel C. and Roberts, Jeanette and Clarke, James S.},
	year = {2025},
}

@article{Chadwick2025,
  title = {Short two-qubit pulse sequences for exchange-only spin qubits in two-dimensional layouts},
  author = {Chadwick, Jason D. and Guerreschi, Gian Giacomo and Luthi, Florian and Madzik, Mateusz T. and Mohiyaddin, Fahd A. and Prabhu, Prithviraj and Schmitz, Albert T. and Litteken, Andrew and Premaratne, Shavindra and Bishop, Nathaniel C. and Matsuura, Anne Y. and Clarke, James S.},
  journal = {Phys. Rev. A},
  volume = {111},
  issue = {5},
  pages = {052616},
  numpages = {15},
  year = {2025},
  month = {May},
  publisher = {American Physical Society},
  doi = {10.1103/PhysRevA.111.052616},
  url = {https://link.aps.org/doi/10.1103/PhysRevA.111.052616}
}

@article{fowler_QFT, 
  author = {Fowler, A. G. and Devitt, S. J. and Hollenberg, L. C. L.},
  title = {Implementation of Shor's algorithm on a linear nearest neighbour qubit array},
  year = {2004},
  issue_date = {July 2004},
  publisher = {Rinton Press, Incorporated},
  address = {Paramus, NJ},
  volume = {4},
  number = {4},
  issn = {1533-7146},
  journal = {Quantum Info. Comput.},
  month = jul, pages = {237–251},
  numpages = {15}
}

@article{weinstein_universal_2023,
	title = {Universal logic with encoded spin qubits in silicon},
	volume = {615},
	issn = {1476-4687},
	url = {https://doi.org/10.1038/s41586-023-05777-3},
	doi = {10.1038/s41586-023-05777-3},
	pages = {817--822},
	number = {7954},
	journal = {Nature},
	author = {Weinstein, Aaron J. and Reed, Matthew D. and Jones, Aaron M. and Andrews, Reed W. and Barnes, David and Blumoff, Jacob Z. and Euliss, Larken E. and Eng, Kevin and Fong, Bryan H. and Ha, Sieu D. and Hulbert, Daniel R. and Jackson, Clayton A. C. and Jura, Michael and Keating, Tyler E. and Kerckhoff, Joseph and Kiselev, Andrey A. and Matten, Justine and Sabbir, Golam and Smith, Aaron and Wright, Jeffrey and Rakher, Matthew T. and Ladd, Thaddeus D. and Borselli, Matthew G.},
	year = {2023},
}

@article{zeuch_efficient_2020,
  title = {Efficient two-qubit pulse sequences beyond CNOT},
  author = {Zeuch, D. and Bonesteel, N. E.},
  journal = {Phys. Rev. B},
  volume = {102},
  issue = {7},
  pages = {075311},
  numpages = {19},
  year = {2020},
  month = {Aug},
  publisher = {American Physical Society},
  doi = {10.1103/PhysRevB.102.075311},
  url = {https://link.aps.org/doi/10.1103/PhysRevB.102.075311}
}

@article{PhysRevLett.85.1758,
  title = {Universal Fault-Tolerant Quantum Computation on Decoherence-Free Subspaces},
  author = {Bacon, D. and Kempe, J. and Lidar, D. A. and Whaley, K. B.},
  journal = {Phys. Rev. Lett.},
  volume = {85},
  issue = {8},
  pages = {1758--1761},
  numpages = {0},
  year = {2000},
  month = {Aug},
  publisher = {American Physical Society},
  doi = {10.1103/PhysRevLett.85.1758},
  url = {https://link.aps.org/doi/10.1103/PhysRevLett.85.1758}
}

@article{PhysRevLett.89.147902,
  title = {Universal Quantum Computation with Spin-$1/2$ Pairs and Heisenberg Exchange},
  author = {Levy, Jeremy},
  journal = {Phys. Rev. Lett.},
  volume = {89},
  issue = {14},
  pages = {147902},
  numpages = {3},
  year = {2002},
  month = {Sep},
  publisher = {American Physical Society},
  doi = {10.1103/PhysRevLett.89.147902},
  url = {https://link.aps.org/doi/10.1103/PhysRevLett.89.147902}
}

@article{PhysRevB.95.241303,
  title = {Exchange-only singlet-only spin qubit},
  author = {Sala, Arnau and Danon, Jeroen},
  journal = {Phys. Rev. B},
  volume = {95},
  issue = {24},
  pages = {241303(R)},
  numpages = {5},
  year = {2017},
  month = {Jun},
  publisher = {American Physical Society},
  doi = {10.1103/PhysRevB.95.241303},
  url = {https://link.aps.org/doi/10.1103/PhysRevB.95.241303}
}

@article{DiVincenzo2000,
	author = {DiVincenzo, D. P. and Bacon, D. and Kempe, J. and Burkard, G. and Whaley, K. B.},
	date = {2000/11/01},
	doi = {10.1038/35042541},
	isbn = {1476-4687},
	journal = {Nature},
	number = {6810},
	pages = {339--342},
	title = {Universal quantum computation with the exchange interaction},
	url = {https://doi.org/10.1038/35042541},
	volume = {408},
	year = {2000},
}

@article{FongWandzura,
	author = {Bryan H. Fong and Stephen M. Wandzura},
	journal = {Quantum.Inf.Comput. 11 (2011)},
	number = {11},
	pages = {1003-1018},
	title = {Universal Quantum Computation and Leakage Reduction in the 3-Qubit Decoherence Free Subsystem},
	year = {2011},
}

@article{Russ2017,
	author = {M. Russ and Burkard, G.},
	date = {2017/08/24},
	doi = {10.1088/1361-648X/aa761f},
	journal = {J. Phys.: Condens. Matter},
	pages = {393001},
	title = {three-electron spin qubits},
	url = {https://iopscience.iop.org/article/10.1088/1361-648X/aa761f},
	volume = {29},
	year = {2017},
}

@article{heinz_fast_2025,
  title = {Fast Quantum Gates for Exchange-Only Qubits Using Simultaneous Exchange Pulses},
  author = {Heinz, Irina and Borjans, Felix and Curry, Matthew J. and Kotlyar, Roza and Luthi, Florian and M\k{a}dzik, Mateusz T. and Mohiyaddin, Fahd A. and Bishop, Nathaniel and Burkard, Guido},
  journal = {PRX Quantum},
  volume = {6},
  issue = {3},
  pages = {030353},
  numpages = {14},
  year = {2025},
  month = {Sep},
  publisher = {American Physical Society},
  doi = {10.1103/njq3-fcdd},
  url = {https://link.aps.org/doi/10.1103/njq3-fcdd}
}

@article{Ha_flexible_2022,
	author = {Ha, Wonill and Ha, Sieu D. and Choi, Maxwell D. and Tang, Yan and Schmitz, Adele E. and Levendorf, Mark P. and Lee, Kangmu and Chappell, James M. and Adams, Tower S. and Hulbert, Daniel R. and Acuna, Edwin and Noah, Ramsey S. and Matten, Justine W. and Jura, Michael P. and Wright, Jeffrey A. and Rakher, Matthew T. and Borselli, Matthew G.},
	date = {2022/02/09},
	doi = {10.1021/acs.nanolett.1c03026},
	isbn = {1530-6984},
	journal = {Nano Lett.},
	month = {02},
	number = {3},
	pages = {1443--1448},
	publisher = {American Chemical Society},
	title = {A Flexible Design Platform for Si/SiGe Exchange-Only Qubits with Low Disorder},
	url = {https://doi.org/10.1021/acs.nanolett.1c03026},
	volume = {22},
	year = {2022},
}

@article{Ha_two_dimensional_2025,
  title = {Two-Dimensional $\mathrm{Si}$ Spin Qubit Arrays with Multilevel Interconnects},
  author = {Ha, Sieu D. and Acuna, Edwin and Raach, Kate and Bloom, Zachery T. and Brecht, Teresa L. and Chappell, James M. and Choi, Maxwell D. and Christensen, Justin E. and Counts, Ian T. and Daprano, Dominic and Dodson, J.P. and Eng, Kevin and Fialkow, David J. and Garcia, Christina A. C. and Ha, Wonill and Harris, Thomas R. B. and holman, nathan and Khalaf, Isaac and Matten, Justine W. and Peterson, Christi A. and Plesha, Clifford E. and Ruiz, Matthew J. and Smith, Aaron and Thomas, Bryan J. and Whiteley, Samuel J. and Ladd, Thaddeus D. and Jura, Michael P. and Rakher, Matthew T. and Borselli, Matthew G.},
  journal = {PRX Quantum},
  volume = {6},
  issue = {3},
  pages = {030327},
  numpages = {7},
  year = {2025},
  month = {Aug},
  publisher = {American Physical Society},
  doi = {10.1103/sgn1-1t2d},
  url = {https://link.aps.org/doi/10.1103/sgn1-1t2d}
}

@misc{Chadwick2025_QEC,
  author = {Jason D. Chadwick and Willers Yang and Joshua Viszlai and Frederic T. Chong},
  title = {A manufacturable surface code architecture for spin qubits with fast transversal logic},
  year={2025},
  eprint={2512.07131},
  archivePrefix={arXiv},
  primaryClass={quant-ph}
}

@misc{Hoffmann2026,
  author = {Silas Hoffman and Edward H. Chen and Matthew Brooks and Stephen Carr and Daniel Volya and Alan Tran and Tyler Keating and Thaddeus D. Ladd and Charles Tahan},
  title = {Spin singlets are useful},
  year={2026},
  eprint={2607.06672},
  archivePrefix={arXiv},
  primaryClass={quant-ph}
}

@misc{Broz2025,
  author = {Joseph D. Broz and Jesse C. Hoke and Edwin Acuna and Jason R. Petta},
  title = {Demonstration of an always-on exchange-only spin qubit},
  year={2025},
  eprint={2508.01033},
  archivePrefix={arXiv},
  primaryClass={quant-ph}
}

@article{Lechner2020,
  author={Lechner, Wolfgang},
  journal={IEEE Transactions on Quantum Engineering}, 
  title={Quantum Approximate Optimization With Parallelizable Gates}, 
  year={2020},
  volume={1},
  number={},
  pages={1-6},
  doi={10.1109/TQE.2020.3034798}
}

@article{Fellner2022applications,
  title = {Applications of universal parity quantum computation},
  author = {Fellner, Michael and Messinger, Anette and Ender, Kilian and Lechner, Wolfgang},
  journal = {Phys. Rev. A},
  volume = {106},
  issue = {4},
  pages = {042442},
  numpages = {8},
  year = {2022},
  month = {Oct},
  publisher = {American Physical Society},
  doi = {10.1103/PhysRevA.106.042442},
  url = {https://link.aps.org/doi/10.1103/PhysRevA.106.042442}
}

@article{ginzel_scalable_2024,
  title = {Scalable parity architecture with a shuttling-based spin qubit processor},
  author = {Ginzel, Florian and Fellner, Michael and Ertler, Christian and Schreiber, Lars R. and Bluhm, Hendrik and Lechner, Wolfgang},
  journal = {Phys. Rev. B},
  volume = {110},
  issue = {7},
  pages = {075302},
  numpages = {16},
  year = {2024},
  month = {Aug},
  publisher = {American Physical Society},
  doi = {10.1103/PhysRevB.110.075302},
  url = {https://link.aps.org/doi/10.1103/PhysRevB.110.075302}
}

@article{replacement_type,
  title = {Replacement-type quantum gates},
  author = {Ginzel, Florian and Kazemi, Javad and Torggler, Valentin and Lechner, Wolfgang},
  journal = {Phys. Rev. A},
  volume = {113},
  issue = {2},
  pages = {022621},
  numpages = {21},
  year = {2026},
  month = {Feb},
  publisher = {American Physical Society},
  doi = {10.1103/frwk-cb6n},
  url = {https://link.aps.org/doi/10.1103/frwk-cb6n}
}

@misc{crooks_performance_2018,
  author = {Crooks, Gavin E.},
  title = {Performance of the Quantum Approximate Optimization Algorithm on the Maximum Cut Problem},
  year={2018},
  eprint={1811.08419},
  archivePrefix={arXiv},
  primaryClass={quant-ph}
}

@article{weidenfeller_scaling_2022,
	title = {Scaling of the quantum approximate optimization algorithm on superconducting qubit based hardware},
	volume = {6},
	url = {https://quantum-journal.org/papers/q-2022-12-07-870/},
	doi = {10.22331/q-2022-12-07-870},
	pages = {870},
	journal = {Quantum},
	publisher = {Verein zur F\"orderung des Open Access Publizierens in den Quantenwissenschaften},
	author = {Weidenfeller, Johannes and Valor, Lucia C. and Gacon, Julien and Tornow, Caroline and Bello, Luciano and Woerner, Stefan and Egger, Daniel J.},
	year = {2022},
	date = {2022-12-07},
	langid = {british},
}

@article{swapless,
  title = {SWAP-less implementation of quantum algorithms},
  author = {Klaver, Berend and Rombouts, Stefan M. A. and Fellner, Michael and Messinger, Anette and Ender, Kilian and Ludwig, Katharina and Lechner, Wolfgang},
  journal = {Phys. Rev. A},
  volume = {113},
  issue = {1},
  pages = {012443},
  numpages = {10},
  year = {2026},
  month = {Jan},
  publisher = {American Physical Society},
  doi = {10.1103/2wzk-fnhx},
  url = {https://link.aps.org/doi/10.1103/2wzk-fnhx}
}

@misc{runtime_paper,
  author = {Federico Domínguez and Michael Fellner and Berend Klaver and Stefan Rombouts and Christian Ertler and Wolfgang Lechner},
  title = {Runtime Reduction in Linear Quantum Charge-Coupled Devices using the Parity Flow Formalism},
  year={2024},
  eprint={2410.16382},
  archivePrefix={arXiv},
  primaryClass={quant-ph}
}

@misc{twine,
      title={Connectivity-aware Synthesis of Quantum Algorithms}, 
      author={Florian Dreier and Christoph Fleckenstein and Gregor Aigner and Michael Fellner and Reinhard Stahn and Martin Lanthaler and Wolfgang Lechner},
      year={2025},
      eprint={2501.14020},
      archivePrefix={arXiv},
      primaryClass={quant-ph}
}

@misc{twine_demo,
      title={Demonstrating Record Fidelity for the Quantum Fourier Transform}, 
      author={Philipp Aumann and Michael Fellner and David Alber and Max Cykiert and Christoph Fleckenstein and Roeland ter Hoeven and Leo Stenzel and Riccardo and Valencia-Tortora and Wolfgang Lechner},
      year={2026},
      eprint={2604.12465},
      archivePrefix={arXiv},
      primaryClass={quant-ph}
}

@article{flow,
  title = {Parity flow formalism: Tracking quantum information throughout computation},
  author = {Klaver, Berend and Ludwig, Katharina and Messinger, Anette and Rombouts, Stefan M. A. and Fellner, Michael and Ender, Kilian and Lechner, Wolfgang},
  journal = {Phys. Rev. Res.},
  volume = {8},
  issue = {1},
  pages = {013095},
  numpages = {33},
  year = {2026},
  month = {Jan},
  publisher = {American Physical Society},
  doi = {10.1103/6xlb-l92j},
  url = {https://link.aps.org/doi/10.1103/6xlb-l92j}
}

@misc{aigner_fermion_2026,
  author = {Gregor Aigner and Berend Klaver and Martin Lanthaler and Wolfgang Lechner},
  title = {Fermion lattices can be simulated by same-size qubit lattices with $\mathcal{O}(1)$  interaction overhead},
  year={2026},
  eprint={2605.12600},
  archivePrefix={arXiv},
  primaryClass={quant-ph},
}

@article{burkard_semiconductor_2023,
  title = {Semiconductor spin qubits},
  author = {Burkard, Guido and Ladd, Thaddeus D. and Pan, Andrew and Nichol, John M. and Petta, Jason R.},
  journal = {Rev. Mod. Phys.},
  volume = {95},
  issue = {2},
  pages = {025003},
  numpages = {58},
  year = {2023},
  month = {Jun},
  publisher = {American Physical Society},
  doi = {10.1103/RevModPhys.95.025003},
  url = {https://link.aps.org/doi/10.1103/RevModPhys.95.025003}
}

@misc{MontanezBarrera2025,
      title={Optimizing {QAOA} circuit transpilation with parity twine and {SWAP} network encodings}, 
      author={J. A. Montanez-Barrera and Yanjun Ji and Michael R. von Spakovsky and David E. Bernal Neira and Kristel Michielsen},
      year={2025},
      eprint={2505.17944},
      archivePrefix={arXiv},
      primaryClass={quant-ph}
}

@article{PhysRevA.54.139,
  title = {Approximate quantum Fourier transform and decoherence},
  author = {Barenco, Adriano and Ekert, Artur and Suominen, Kalle-Antti and T\"orm\"a, P\"aivi},
  journal = {Phys. Rev. A},
  volume = {54},
  issue = {1},
  pages = {139--146},
  numpages = {0},
  year = {1996},
  month = {Jul},
  publisher = {American Physical Society},
  doi = {10.1103/PhysRevA.54.139},
  url = {https://link.aps.org/doi/10.1103/PhysRevA.54.139}
}

@misc{Maskara2025,
      title={Fast simulation of fermions with reconfigurable qubits}, 
      author={N. Maskara and M. Kalinowski and D. Gonzalez-Cuadra and  M. D. Lukin},
      year={2025},
      eprint={2509.08898},
      archivePrefix={arXiv},
      primaryClass={quant-ph}
}

@misc{Constantinides2025,
      title={Low-depth fermion routing without ancillas}, 
      author={N. Constantinides and J. Yu and D. Devulapalli and A. Fahimniya and  L. Schaeffer and A. M. Childs and M. J. Gullans and A. Schuckert and A. V. Gorshko},
      year={2025},
      eprint={2510.05099},
      archivePrefix={arXiv},
      primaryClass={quant-ph}
}

@article{PhysRevLett.120.110501,
  title = {Quantum Simulation of Electronic Structure with Linear Depth and Connectivity},
  author = {Kivlichan, Ian D. and McClean, Jarrod and Wiebe, Nathan and Gidney, Craig and Aspuru-Guzik, Al\'an and Chan, Garnet Kin-Lic and Babbush, Ryan},
  journal = {Phys. Rev. Lett.},
  volume = {120},
  issue = {11},
  pages = {110501},
  numpages = {6},
  year = {2018},
  month = {Mar},
  publisher = {American Physical Society},
  doi = {10.1103/PhysRevLett.120.110501},
  url = {https://link.aps.org/doi/10.1103/PhysRevLett.120.110501}
}

@article{PRXQuantum.5.030323,
  title = {Measuring the Loschmidt Amplitude for Finite-Energy Properties of the Fermi-Hubbard Model on an Ion-Trap Quantum Computer},
  author = {H\'emery, K\'evin and Ghanem, Khaldoon and Crane, Eleanor and Campbell, Sara L. and Dreiling, Joan M. and Figgatt, Caroline and Foltz, Cameron and Gaebler, John P. and Johansen, Jacob and Mills, Michael and Moses, Steven A. and Pino, Juan M. and Ransford, Anthony and Rowe, Mary and Siegfried, Peter and Stutz, Russell P. and Dreyer, Henrik and Schuckert, Alexander and Nigmatullin, Ramil},
  journal = {PRX Quantum},
  volume = {5},
  issue = {3},
  pages = {030323},
  numpages = {20},
  year = {2024},
  month = {Aug},
  publisher = {American Physical Society},
  doi = {10.1103/PRXQuantum.5.030323},
  url = {https://link.aps.org/doi/10.1103/PRXQuantum.5.030323}
}

@misc{Alam2025,
  author = {Faisal Alam and Jan Lukas Bosse and Ieva \v{C}epait\.e and Adrian Chapman and Laura Clinton and Marcos Crichigno and Elizabeth Crosson and Toby Cubitt and Charles Derby and Oliver Dowinton and Paul K. Faehrmann and Steve Flammia and Brian Flynn and Filippo Maria Gambetta and Ra\'ul Garc\'ia-Patr\'on and Max Hunter-Gordon and Glenn Jones and Abhishek Khedkar and Joel Klassen and Michael Kreshchuk and Edward Harry McMullan and Lana Mineh and Ashley Montanaro and Caterina Mora and John J. L. Morton and Dhrumil Patel and Pete Rolph and Raul A. Santos and James R. Seddon and Evan Sheridan and Wilfrid Somogyi and Marika Svensson and Niam Vaishnav and Sabrina Yue Wang and Gethin Wright},
  title = {Programmable digital quantum simulation of 2D Fermi-Hubbard dynamics using 72 superconducting qubits},
  year={2025},
  eprint={2510.26845},
  archivePrefix={arXiv},
  primaryClass={quant-ph}
}

@article{PhysRevA.79.032316,
  title = {Quantum circuits for strongly correlated quantum systems},
  author = {Verstraete, Frank and Cirac, J. Ignacio and Latorre, Jos\'e I.},
  journal = {Phys. Rev. A},
  volume = {79},
  issue = {3},
  pages = {032316},
  numpages = {5},
  year = {2009},
  month = {Mar},
  publisher = {American Physical Society},
  doi = {10.1103/PhysRevA.79.032316},
  url = {https://link.aps.org/doi/10.1103/PhysRevA.79.032316}
}

@article{PhysRevX.8.011044,
  title = {Low-Depth Quantum Simulation of Materials},
  author = {Babbush, Ryan and Wiebe, Nathan and McClean, Jarrod and McClain, James and Neven, Hartmut and Chan, Garnet Kin-Lic},
  journal = {Phys. Rev. X},
  volume = {8},
  issue = {1},
  pages = {011044},
  numpages = {40},
  year = {2018},
  month = {Mar},
  publisher = {American Physical Society},
  doi = {10.1103/PhysRevX.8.011044},
  url = {https://link.aps.org/doi/10.1103/PhysRevX.8.011044}
}

@article{DumoulinStuyck2026,
	author = {Dumoulin Stuyck, Nard and Saraiva, Andre and Gilbert, Will and Cifuentes Pardo, Jesus and Li, Ruoyu and Escott, Christopher C. and De Greve, Kristiaan and Voinigescu, Sorin and Reilly, David J. and Dzurak, Andrew S.},
	date = {2026/05/01},
	doi = {10.1038/s44287-026-00283-w},
	isbn = {2948-1201},
	journal = {Nat. Rev. El. Eng.},
	number = {5},
	pages = {300--315},
	title = {{CMOS} compatibility of semiconductor spin qubits},
	url = {https://doi.org/10.1038/s44287-026-00283-w},
	volume = {3},
	year = {2026},
}

@misc{McIntyre2026,
  author = {Z. M. McIntyre, Abhikbrata Sarkar, Daniel Loss},
  title = {Theory of spin qubits and the path to scalability},
  year={2026},
  eprint={2604.13644},
  archivePrefix={arXiv},
  primaryClass={quant-ph}
}

@misc{SqDRIFT,
  author = {Samuele Piccinelli and Alberto Baiardi and Stefano Barison and Max Rossmannek and Almudena Carrera Vazquez and Francesco Tacchino and Stefano Mensa and Edoardo Altamura and Ali Alavi and Mario Motta and Javier Robledo-Moreno and William Kirby and Kunal Sharma and Antonio Mezzacapo and Ivano Tavernelli},
  title = {Quantum chemistry with provable convergence via randomized sample-based Krylov quantum diagonalization},
  year={2025},
  eprint={2508.02578},
  archivePrefix={arXiv},
  primaryClass={quant-ph}
}

@article{doi:10.1073/pnas.1619152114,
author = {Markus Reihe and Nathan Wiebe and Krysta M. Svore and Dave Wecker and Matthias Troyer },
title = {Elucidating reaction mechanisms on quantum computers},
journal = {Proc. Natl. Acad. Sci.},
volume = {114},
number = {29},
pages = {7555-7560},
year = {2017},
doi = {10.1073/pnas.1619152114},
URL = {https://www.pnas.org/doi/abs/10.1073/pnas.1619152114},
}

@article{PhysRevResearch.4.033043,
  title = {Long-range electron-electron interactions in quantum dot systems and applications in quantum chemistry},
  author = {Kn\"orzer, J. and van Diepen, C. J. and Hsiao, T.-K. and Giedke, G. and Mukhopadhyay, U. and Reichl, C. and Wegscheider, W. and Cirac, J. I. and Vandersypen, L. M. K.},
  journal = {Phys. Rev. Res.},
  volume = {4},
  issue = {3},
  pages = {033043},
  numpages = {10},
  year = {2022},
  month = {Jul},
  publisher = {American Physical Society},
  doi = {10.1103/PhysRevResearch.4.033043},
  url = {https://link.aps.org/doi/10.1103/PhysRevResearch.4.033043}
}

@article{Dehollain2020,
	author = {Dehollain, J. P. and Mukhopadhyay, U. and Michal, V. P. and Wang, Y. and Wunsch, B. and Reichl, C. and Wegscheider, W. and Rudner, M. S. and Demler, E. and Vandersypen, L. M. K.},
	date = {2020/03/01},
	doi = {10.1038/s41586-020-2051-0},
	isbn = {1476-4687},
	journal = {Nature},
	number = {7800},
	pages = {528--533},
	title = {Nagaoka ferromagnetism observed in a quantum dot plaquette},
	url = {https://doi.org/10.1038/s41586-020-2051-0},
	volume = {579},
	year = {2020},
}

@article{Kiczynski2022,
	author = {Kiczynski, M. and Gorman, S. K. and Geng, H. and Donnelly, M. B. and Chung, Y. and He, Y. and Keizer, J. G. and Simmons, M. Y.},
	date = {2022/06/01},
	doi = {10.1038/s41586-022-04706-0},
	isbn = {1476-4687},
	journal = {Nature},
	number = {7915},
	pages = {694--699},
	title = {Engineering topological states in atom-based semiconductor quantum dots},
	url = {https://doi.org/10.1038/s41586-022-04706-0},
	volume = {606},
	year = {2022},
}

@misc{Morozova2026,
  author = {Elizaveta Morozova and Xin Zhang and Utso Bhattacharya and Pablo Cova Fari{\~n}a and Daniel Jirovec and Alexander Nico-Katz and Stefan D. Oosterhout and Sougato Bose and Giordano Scappucci and Menno Veldhorst and Eugene Demler and Lieven M. K. Vandersypen},
  title = {Observation of magnetic quantum phase crossovers in a semiconductor spin ladder},
  year={2026},
  eprint={2608.17789},
  archivePrefix={arXiv},
  primaryClass={cond-mat.mes-hal}
}

@article{Struck2024,
	author = {Struck, Tom and Volmer, Mats and Visser, Lino and Offermann, Tobias and Xue, Ran and Tu, Jhih-Sian and Trellenkamp, Stefan and Cywi{\'n}ski, {\L}ukasz and Bluhm, Hendrik and Schreiber, Lars R.},
	date = {2024/02/13},
	doi = {10.1038/s41467-024-45583-7},
	isbn = {2041-1723},
	journal = {Nat. Commun.},
	number = {1},
	pages = {1325},
	title = {Spin-EPR-pair separation by conveyor-mode single electron shuttling in Si/SiGe},
	url = {https://doi.org/10.1038/s41467-024-45583-7},
	volume = {15},
	year = {2024}
}

@article{DeSmet2025,
	author = {De Smet, Maxim and Matsumoto, Yuta and Zwerver, Anne-Marije J. and Tryputen, Larysa and de Snoo, Sander L. and Amitonov, Sergey V. and Katiraee-Far, Sam R. and Sammak, Amir and Samkharadze, Nodar and G{\"u}l, {\"O}nder and Wasserman, Rick N. M. and Greplov{\'a}, Eli{\v s}ka and Rimbach-Russ, Maximilian and Scappucci, Giordano and Vandersypen, Lieven M. K.},
	date = {2025/07/01},
	doi = {10.1038/s41565-025-01920-5},
	isbn = {1748-3395},
	journal = {Nat. Nanotechnol.},
	number = {7},
	pages = {866--872},
	title = {High-fidelity single-spin shuttling in silicon},
	url = {https://doi.org/10.1038/s41565-025-01920-5},
	volume = {20},
	year = {2025}
}

@misc{Unseth_2026_parity,
  author = {Brennan Undseth and Nicola Meggiato and Yi-Hsien Wu and Sam R. Katiraee-Far and Larysa Tryputen and Sander L. de Snoo and Davide Degli Esposti and Giordano Scappucci and Eli\v{s}ka Greplov\'a and Lieven M. K. Vandersypen},
  title = {Weight-four parity checks with silicon spin qubits},
  year={2026},
  eprint={2601.23267},
  archivePrefix={arXiv},
  primaryClass={cond-mat.mes-hall}
}

@article{BRAVYI2002210,
title = {Fermionic Quantum Computation},
journal = {Annals of Physics},
volume = {298},
number = {1},
pages = {210-226},
year = {2002},
issn = {0003-4916},
doi = {https://doi.org/10.1006/aphy.2002.6254},
url = {https://www.sciencedirect.com/science/article/pii/S0003491602962548},
author = {Sergey B. Bravyi and Alexei Yu. Kitaev},
}

@article{Verstraete_2005,
doi = {10.1088/1742-5468/2005/09/P09012},
url = {https://doi.org/10.1088/1742-5468/2005/09/P09012},
year = {2005},
month = {sep},
publisher = {},
volume = {2005},
number = {09},
pages = {P09012},
author = {Verstraete, F and Cirac, J I},
title = {Mapping local Hamiltonians of fermions to local Hamiltonians of spins},
journal = {Journal of Statistical Mechanics: Theory and Experiment},
}

@article{PhysRevB.104.035118,
  title = {Compact fermion to qubit mappings},
  author = {Derby, Charles and Klassen, Joel and Bausch, Johannes and Cubitt, Toby},
  journal = {Phys. Rev. B},
  volume = {104},
  issue = {3},
  pages = {035118},
  numpages = {12},
  year = {2021},
  month = {Jul},
  publisher = {American Physical Society},
  doi = {10.1103/PhysRevB.104.035118},
  url = {https://link.aps.org/doi/10.1103/PhysRevB.104.035118}
}

@misc{Dijkema2026,
  author = {Jurgen J. Dijkema and Xin Zhang and Achilleas Bardakas and Daniel Bouman and Alice Cuzzocrea and David van Driel and Davide Girardi and Lucas E.A. Stehouwer and Giordano Scappucci and Anne-Marije J. Zwerver and Nico W. Hendrickx},
  title = {Simultaneous operation of an 18-qubit modular array in germanium},
  year={2026},
  eprint={2604.01063},
  archivePrefix={arXiv},
  primaryClass={cond-mat.mes-hall}
}

@article{Li2026,
	author = {Li, R. and Levajac, V. and Godfrin, C. and Kubicek, S. and Simion, G. and Raes, B. and Beyne, S. and Fattal, I. and Loenders, A. and De Roeck, W. and Mongillo, M. and Wan, D. and De Greve, K.},
	date = {2026/04/06},
	doi = {10.1038/s41598-026-42575-z},
	journal = {Scientific Reports},
	number = {1},
	pages = {16526},
	title = {A tri-linear quantum dot architecture for semiconductor spin qubits},
	url = {https://doi.org/10.1038/s41598-026-42575-z},
	volume = {16},
	year = {2026},}

@article{Preskill2018quantumcomputingin,
  doi = {10.22331/q-2018-08-06-79},
  url = {https://doi.org/10.22331/q-2018-08-06-79},
  title = {Quantum {C}omputing in the {NISQ} era and beyond},
  author = {Preskill, John},
  journal = {{Quantum}},
  issn = {2521-327X},
  publisher = {{Verein zur F{\"{o}}rderung des Open Access Publizierens in den Quantenwissenschaften}},
  volume = {2},
  pages = {79},
  month = aug,
  year = {2018}
}

@book{NielsenChuang,
    author = {M. A. Nielsen and I. L. Chuang},
    title = {Quantum Computation and Quantum Information},
    publisher = {Cambrige University Press},
    address = {Cambridge},
    year = {2000}
}

@article{Preskill1998,
  doi = {10.1098/rspa.1998.0167},
  url = {http://doi.org/10.1098/rspa.1998.0167},
  title = {Reliable quantum computers},
  author = {Preskill, John},
  journal = {Proc. R. Soc. Lond. A.},
  volume = {454},
  pages = {385-410},
  year = {1998}
}

@article{PhysRevA.86.032324,
  title = {Surface codes: Towards practical large-scale quantum computation},
  author = {Fowler, Austin G. and Mariantoni, Matteo and Martinis, John M. and Cleland, Andrew N.},
  journal = {Phys. Rev. A},
  volume = {86},
  issue = {3},
  pages = {032324},
  numpages = {48},
  year = {2012},
  month = {Sep},
  publisher = {American Physical Society},
  doi = {10.1103/PhysRevA.86.032324},
  url = {https://link.aps.org/doi/10.1103/PhysRevA.86.032324}
}

@article{Campbell_2017_fault_tolerant,
	author = {Campbell, Earl T. and Terhal, Barbara M. and Vuillot, Christophe},
	date = {2017/09/01},
	doi = {10.1038/nature23460},
	isbn = {1476-4687},
	journal = {Nature},
	number = {7671},
	pages = {172--179},
	title = {Roads towards fault-tolerant universal quantum computation},
	url = {https://doi.org/10.1038/nature23460},
	volume = {549},
	year = {2017},
}

@article{Litinski2019gameofsurfacecodes,
  doi = {10.22331/q-2019-03-05-128},
  url = {https://doi.org/10.22331/q-2019-03-05-128},
  title = {A {G}ame of {S}urface {C}odes: {L}arge-{S}cale {Q}uantum {C}omputing with {L}attice {S}urgery},
  author = {Litinski, Daniel},
  journal = {{Quantum}},
  issn = {2521-327X},
  publisher = {{Verein zur F{\"{o}}rderung des Open Access Publizierens in den Quantenwissenschaften}},
  volume = {3},
  pages = {128},
  month = mar,
  year = {2019}
}

@article{Litinski2018latticesurgery,
  doi = {10.22331/q-2018-05-04-62},
  url = {https://doi.org/10.22331/q-2018-05-04-62},
  title = {Lattice {S}urgery with a {T}wist: {S}implifying {C}lifford {G}ates of {S}urface {C}odes},
  author = {Litinski, Daniel and Oppen, Felix von},
  journal = {{Quantum}},
  issn = {2521-327X},
  publisher = {{Verein zur F{\"{o}}rderung des Open Access Publizierens in den Quantenwissenschaften}},
  volume = {2},
  pages = {62},
  month = may,
  year = {2018}
}

@article{Google2021,
	author = {Chen, Zijun and Satzinger, Kevin J. and Atalaya, Juan and Korotkov, Alexander N. and Dunsworth, Andrew and Sank, Daniel and Quintana, Chris and McEwen, Matt and Barends, Rami and Klimov, Paul V. and Hong, Sabrina and Jones, Cody and Petukhov, Andre and Kafri, Dvir and Demura, Sean and Burkett, Brian and Gidney, Craig and Fowler, Austin G. and Paler, Alexandru and Putterman, Harald and Aleiner, Igor and Arute, Frank and Arya, Kunal and Babbush, Ryan and Bardin, Joseph C. and Bengtsson, Andreas and Bourassa, Alexandre and Broughton, Michael and Buckley, Bob B. and Buell, David A. and Bushnell, Nicholas and Chiaro, Benjamin and Collins, Roberto and Courtney, William and Derk, Alan R. and Eppens, Daniel and Erickson, Catherine and Farhi, Edward and Foxen, Brooks and Giustina, Marissa and Greene, Ami and Gross, Jonathan A. and Harrigan, Matthew P. and Harrington, Sean D. and Hilton, Jeremy and Ho, Alan and Huang, Trent and Huggins, William J. and Ioffe, L. B. and Isakov, Sergei V. and Jeffrey, Evan and Jiang, Zhang and Kechedzhi, Kostyantyn and Kim, Seon and Kitaev, Alexei and Kostritsa, Fedor and Landhuis, David and Laptev, Pavel and Lucero, Erik and Martin, Orion and McClean, Jarrod R. and McCourt, Trevor and Mi, Xiao and Miao, Kevin C. and Mohseni, Masoud and Montazeri, Shirin and Mruczkiewicz, Wojciech and Mutus, Josh and Naaman, Ofer and Neeley, Matthew and Neill, Charles and Newman, Michael and Niu, Murphy Yuezhen and O'Brien, Thomas E. and Opremcak, Alex and Ostby, Eric and Pat{\'o}, B{\'a}lint and Redd, Nicholas and Roushan, Pedram and Rubin, Nicholas C. and Shvarts, Vladimir and Strain, Doug and Szalay, Marco and Trevithick, Matthew D. and Villalonga, Benjamin and White, Theodore and Yao, Z. Jamie and Yeh, Ping and Yoo, Juhwan and Zalcman, Adam and Neven, Hartmut and Boixo, Sergio and Smelyanskiy, Vadim and Chen, Yu and Megrant, Anthony and Kelly, Julian and {Google Quantum AI}},
	date = {2021/07/01},
	doi = {10.1038/s41586-021-03588-y},
	isbn = {1476-4687},
	journal = {Nature},
	number = {7867},
	pages = {383--387},
	title = {Exponential suppression of bit or phase errors with cyclic error correction},
	url = {https://doi.org/10.1038/s41586-021-03588-y},
	volume = {595},
	year = {2021},}

@article{Google2025,
	author = {Acharya, Rajeev and Abanin, Dmitry A. and Aghababaie-Beni, Laleh and Aleiner, Igor and Andersen, Trond I. and Ansmann, Markus and Arute, Frank and Arya, Kunal and Asfaw, Abraham and Astrakhantsev, Nikita and Atalaya, Juan and Babbush, Ryan and Bacon, Dave and Ballard, Brian and Bardin, Joseph C. and Bausch, Johannes and Bengtsson, Andreas and Bilmes, Alexander and Blackwell, Sam and Boixo, Sergio and Bortoli, Gina and Bourassa, Alexandre and Bovaird, Jenna and Brill, Leon and Broughton, Michael and Browne, David A. and Buchea, Brett and Buckley, Bob B. and Buell, David A. and Burger, Tim and Burkett, Brian and Bushnell, Nicholas and Cabrera, Anthony and Campero, Juan and Chang, Hung-Shen and Chen, Yu and Chen, Zijun and Chiaro, Ben and Chik, Desmond and Chou, Charina and Claes, Jahan and Cleland, Agnetta Y. and Cogan, Josh and Collins, Roberto and Conner, Paul and Courtney, William and Crook, Alexander L. and Curtin, Ben and Das, Sayan and Davies, Alex and De Lorenzo, Laura and Debroy, Dripto M. and Demura, Sean and Devoret, Michel and Di Paolo, Agustin and Donohoe, Paul and Drozdov, Ilya and Dunsworth, Andrew and Earle, Clint and Edlich, Thomas and Eickbusch, Alec and Elbag, Aviv Moshe and Elzouka, Mahmoud and Erickson, Catherine and Faoro, Lara and Farhi, Edward and Ferreira, Vinicius S. and Burgos, Leslie Flores and Forati, Ebrahim and Fowler, Austin G. and Foxen, Brooks and Ganjam, Suhas and Garcia, Gonzalo and Gasca, Robert and Genois, {\'E}lie and Giang, William and Gidney, Craig and Gilboa, Dar and Gosula, Raja and Dau, Alejandro Grajales and Graumann, Dietrich and Greene, Alex and Gross, Jonathan A. and Habegger, Steve and Hall, John and Hamilton, Michael C. and Hansen, Monica and Harrigan, Matthew P. and Harrington, Sean D. and Heras, Francisco J. H. and Heslin, Stephen and Heu, Paula and Higgott, Oscar and Hill, Gordon and Hilton, Jeremy and Holland, George and Hong, Sabrina and Huang, Hsin-Yuan and Huff, Ashley and Huggins, William J. and Ioffe, Lev B. and Isakov, Sergei V. and Iveland, Justin and Jeffrey, Evan and Jiang, Zhang and Jones, Cody and Jordan, Stephen and Joshi, Chaitali and Juhas, Pavol and Kafri, Dvir and Kang, Hui and Karamlou, Amir H. and Kechedzhi, Kostyantyn and Kelly, Julian and Khaire, Trupti and Khattar, Tanuj and Khezri, Mostafa and Kim, Seon and Klimov, Paul V. and Klots, Andrey R. and Kobrin, Bryce and Kohli, Pushmeet and Korotkov, Alexander N. and Kostritsa, Fedor and Kothari, Robin and Kozlovskii, Borislav and Kreikebaum, John Mark and Kurilovich, Vladislav D. and Lacroix, Nathan and Landhuis, David and Lange-Dei, Tiano and Langley, Brandon W. and Laptev, Pavel and Lau, Kim-Ming and Le Guevel, Lo{\"\i}ck and Ledford, Justin and Lee, Joonho and Lee, Kenny and Lensky, Yuri D. and Leon, Shannon and Lester, Brian J. and Li, Wing Yan and Li, Yin and Lill, Alexander T. and Liu, Wayne and Livingston, William P. and Locharla, Aditya and Lucero, Erik and Lundahl, Daniel and Lunt, Aaron and Madhuk, Sid and Malone, Fionn D. and Maloney, Ashley and Mandr{\`a}, Salvatore and Manyika, James and Martin, Leigh S. and Martin, Orion and Martin, Steven and Maxfield, Cameron and McClean, Jarrod R. and McEwen, Matt and Meeks, Seneca and Megrant, Anthony and Mi, Xiao and Miao, Kevin C. and Mieszala, Amanda and Molavi, Reza and Molina, Sebastian and Montazeri, Shirin and Morvan, Alexis and Movassagh, Ramis and Mruczkiewicz, Wojciech and Naaman, Ofer and Neeley, Matthew and Neill, Charles and Nersisyan, Ani and Neven, Hartmut and Newman, Michael and Ng, Jiun How and Nguyen, Anthony and Nguyen, Murray and Ni, Chia-Hung and Niu, Murphy Yuezhen and O'Brien, Thomas E. and Oliver, William D. and Opremcak, Alex and Ottosson, Kristoffer and Petukhov, Andre and Pizzuto, Alex and Platt, John and Potter, Rebecca and Pritchard, Orion and Pryadko, Leonid P. and Quintana, Chris and Ramachandran, Ganesh and Reagor, Matthew J. and Redding, John and Rhodes, David M. and Roberts, Gabrielle and Rosenberg, Eliott and Rosenfeld, Emma and Roushan, Pedram and Rubin, Nicholas C. and Saei, Negar and Sank, Daniel and Sankaragomathi, Kannan and Satzinger, Kevin J. and Schurkus, Henry F. and Schuster, Christopher and Senior, Andrew W. and Shearn, Michael J. and Shorter, Aaron and Shutty, Noah and Shvarts, Vladimir and Singh, Shraddha and Sivak, Volodymyr and Skruzny, Jindra and Small, Spencer and Smelyanskiy, Vadim and Smith, W. Clarke and Somma, Rolando D. and Springer, Sofia and Sterling, George and Strain, Doug and Suchard, Jordan and Szasz, Aaron and Sztein, Alex and Thor, Douglas and Torres, Alfredo and Torunbalci, M. Mert and Vaishnav, Abeer and Vargas, Justin and Vdovichev, Sergey and Vidal, Guifre and Villalonga, Benjamin and Heidweiller, Catherine Vollgraff and Waltman, Steven and Wang, Shannon X. and Ware, Brayden and Weber, Kate and Weidel, Travis and White, Theodore and Wong, Kristi and Woo, Bryan W. K. and Xing, Cheng and Yao, Z. Jamie and Yeh, Ping and Ying, Bicheng and Yoo, Juhwan and Yosri, Noureldin and Young, Grayson and Zalcman, Adam and Zhang, Yaxing and Zhu, Ningfeng and Zobrist, Nicholas and {Google Quantum AI} and Collaborators},
	date = {2025/02/01},
	doi = {10.1038/s41586-024-08449-y},
	isbn = {1476-4687},
	journal = {Nature},
	number = {8052},
	pages = {920--926},
	title = {Quantum error correction below the surface code threshold},
	url = {https://doi.org/10.1038/s41586-024-08449-y},
	volume = {638},
	year = {2025}
}

@article{Google2023,
	author = {Acharya, Rajeev and Aleiner, Igor and Allen, Richard and Andersen, Trond I. and Ansmann, Markus and Arute, Frank and Arya, Kunal and Asfaw, Abraham and Atalaya, Juan and Babbush, Ryan and Bacon, Dave and Bardin, Joseph C. and Basso, Joao and Bengtsson, Andreas and Boixo, Sergio and Bortoli, Gina and Bourassa, Alexandre and Bovaird, Jenna and Brill, Leon and Broughton, Michael and Buckley, Bob B. and Buell, David A. and Burger, Tim and Burkett, Brian and Bushnell, Nicholas and Chen, Yu and Chen, Zijun and Chiaro, Ben and Cogan, Josh and Collins, Roberto and Conner, Paul and Courtney, William and Crook, Alexander L. and Curtin, Ben and Debroy, Dripto M. and Del Toro Barba, Alexander and Demura, Sean and Dunsworth, Andrew and Eppens, Daniel and Erickson, Catherine and Faoro, Lara and Farhi, Edward and Fatemi, Reza and Flores Burgos, Leslie and Forati, Ebrahim and Fowler, Austin G. and Foxen, Brooks and Giang, William and Gidney, Craig and Gilboa, Dar and Giustina, Marissa and Grajales Dau, Alejandro and Gross, Jonathan A. and Habegger, Steve and Hamilton, Michael C. and Harrigan, Matthew P. and Harrington, Sean D. and Higgott, Oscar and Hilton, Jeremy and Hoffmann, Markus and Hong, Sabrina and Huang, Trent and Huff, Ashley and Huggins, William J. and Ioffe, Lev B. and Isakov, Sergei V. and Iveland, Justin and Jeffrey, Evan and Jiang, Zhang and Jones, Cody and Juhas, Pavol and Kafri, Dvir and Kechedzhi, Kostyantyn and Kelly, Julian and Khattar, Tanuj and Khezri, Mostafa and Kieferov{\'a}, M{\'a}ria and Kim, Seon and Kitaev, Alexei and Klimov, Paul V. and Klots, Andrey R. and Korotkov, Alexander N. and Kostritsa, Fedor and Kreikebaum, John Mark and Landhuis, David and Laptev, Pavel and Lau, Kim-Ming and Laws, Lily and Lee, Joonho and Lee, Kenny and Lester, Brian J. and Lill, Alexander and Liu, Wayne and Locharla, Aditya and Lucero, Erik and Malone, Fionn D. and Marshall, Jeffrey and Martin, Orion and McClean, Jarrod R. and McCourt, Trevor and McEwen, Matt and Megrant, Anthony and Meurer Costa, Bernardo and Mi, Xiao and Miao, Kevin C. and Mohseni, Masoud and Montazeri, Shirin and Morvan, Alexis and Mount, Emily and Mruczkiewicz, Wojciech and Naaman, Ofer and Neeley, Matthew and Neill, Charles and Nersisyan, Ani and Neven, Hartmut and Newman, Michael and Ng, Jiun How and Nguyen, Anthony and Nguyen, Murray and Niu, Murphy Yuezhen and O'Brien, Thomas E. and Opremcak, Alex and Platt, John and Petukhov, Andre and Potter, Rebecca and Pryadko, Leonid P. and Quintana, Chris and Roushan, Pedram and Rubin, Nicholas C. and Saei, Negar and Sank, Daniel and Sankaragomathi, Kannan and Satzinger, Kevin J. and Schurkus, Henry F. and Schuster, Christopher and Shearn, Michael J. and Shorter, Aaron and Shvarts, Vladimir and Skruzny, Jindra and Smelyanskiy, Vadim and Smith, W. Clarke and Sterling, George and Strain, Doug and Szalay, Marco and Torres, Alfredo and Vidal, Guifre and Villalonga, Benjamin and Vollgraff Heidweiller, Catherine and White, Theodore and Xing, Cheng and Yao, Z. Jamie and Yeh, Ping and Yoo, Juhwan and Young, Grayson and Zalcman, Adam and Zhang, Yaxing and Zhu, Ningfeng and {Google Quantum AI}},
	date = {2023/02/01},
	doi = {10.1038/s41586-022-05434-1},
	isbn = {1476-4687},
	journal = {Nature},
	number = {7949},
	pages = {676--681},
	title = {Suppressing quantum errors by scaling a surface code logical qubit},
	url = {https://doi.org/10.1038/s41586-022-05434-1},
	volume = {614},
	year = {2023}
}

@article{Siegel_2026_snakes,
  title = {Quantum Snakes on a Plane: Mobile, Low-Dimensional Logical Qubits on a 2D Surface},
  author = {Siegel, Adam and Cai, Zhenyu and Jnane, Hamza and Koczor, Balint and Pexton, Shaun and Strikis, Armands and Benjamin, Simon},
  journal = {PRX Quantum},
  volume = {7},
  issue = {1},
  pages = {010339},
  numpages = {30},
  year = {2026},
  month = {Feb},
  publisher = {American Physical Society},
  doi = {10.1103/494s-jd8h},
  url = {https://link.aps.org/doi/10.1103/494s-jd8h}
}

@misc{Eggerickx_2026,
  author = {Quinten Eggerickx and Simon C. Benjamin},
  title = {A route to damage tolerance exceeding $10\%$ in shuttling-equipped quantum processors},
  year={2026},
  eprint={2607.22429},
  archivePrefix={arXiv},
  primaryClass={quant-ph}
}

@misc{Siegel_2026_erasure,
  author = {Adam Siegel and Simon C. Benjamin},
  title = {Erasure conversion for singlet-triplet spin qubits enables high-performance shuttling-based quantum error correction},
  year={2026},
  eprint={2601.10461},
  archivePrefix={arXiv},
  primaryClass={quant-ph}
}

@misc{Steinacker_2026_temperature,
  author = {Paul Steinacker and Amanda E. Seedhouse and Nard Dumoulin Stuyck and Tuomo Tanttu and MengKe Feng and Santiago Serrano and Ensar Vahapoglu and Samuel K. Bartee and Philip Y. Mai and Alexis Shaw and Andreas Nickl and Sebastian Pauka and Brendan Harlech-Jones and Juan P. Dehollain and Fay E. Hudson and Kok Wai Chan and Thomas A. Ohki and David Reilly and Christopher C. Escott and Chih Hwan Yang and Wee Han Lim and Arne Laucht and Andre Saraiva and Andrew S. Dzurak and Jared H. Cole},
  title = {Optimal operating temperature for industry-compatible silicon spin quantum computing: colder is not necessarily better},
  year={2026},
  eprint={2607.11846},
  archivePrefix={arXiv},
  primaryClass={quant-ph}
}

@article{PhysRevA.109.032433,
  title = {Tailoring quantum error correction to spin qubits},
  author = {Het\'enyi, Bence and Wootton, James R.},
  journal = {Phys. Rev. A},
  volume = {109},
  issue = {3},
  pages = {032433},
  numpages = {22},
  year = {2024},
  month = {Mar},
  publisher = {American Physical Society},
  doi = {10.1103/PhysRevA.109.032433},
  url = {https://link.aps.org/doi/10.1103/PhysRevA.109.032433}
}

@article{PRXQuantum.5.040328,
  title = {Towards Early Fault Tolerance on a $2\ifmmode\times\else\texttimes\fi{}N$ Array of Qubits Equipped with Shuttling},
  author = {Siegel, Adam and Strikis, Armands and Fogarty, Michael},
  journal = {PRX Quantum},
  volume = {5},
  issue = {4},
  pages = {040328},
  numpages = {22},
  year = {2024},
  month = {Nov},
  publisher = {American Physical Society},
  doi = {10.1103/PRXQuantum.5.040328},
  url = {https://link.aps.org/doi/10.1103/PRXQuantum.5.040328}
}

@article{Dennis2002_topological,
  title = {Topological quantum memory},
  author = {Eric Dennis and Alexei Kitaev and Andrew Landahl and John Preskill},
  journal = {J. Math. Phys.},
  volume = {43},
  issue = {9},
  pages = {4452},
  numpages = {53},
  year = {2002},
  month = {Sep},
  doi = {https://doi.org/10.1063/1.1499754},
  url = {https://pubs.aip.org/aip/jmp/article-abstract/43/9/4452/230976/Topological-quantum-memory}
}

@article{PhysRevA.62.052316,
  title = {Methodology for quantum logic gate construction},
  author = {Zhou, Xinlan and Leung, Debbie W. and Chuang, Isaac L.},
  journal = {Phys. Rev. A},
  volume = {62},
  issue = {5},
  pages = {052316},
  numpages = {12},
  year = {2000},
  month = {Oct},
  publisher = {American Physical Society},
  doi = {10.1103/PhysRevA.62.052316},
  url = {https://link.aps.org/doi/10.1103/PhysRevA.62.052316}
}

@article{Gidney2019efficientmagicstate,
  doi = {10.22331/q-2019-04-30-135},
  url = {https://doi.org/10.22331/q-2019-04-30-135},
  title = {Efficient magic state factories with a catalyzed {$|CCZ\rangle$} to {$2|T\rangle$} transformation},
  author = {Gidney, Craig and Fowler, Austin G.},
  journal = {{Quantum}},
  issn = {2521-327X},
  publisher = {{Verein zur F{\"{o}}rderung des Open Access Publizierens in den Quantenwissenschaften}},
  volume = {3},
  pages = {135},
  month = apr,
  year = {2019}
}

@article{Litinski2019magicstate,
  doi = {10.22331/q-2019-12-02-205},
  url = {https://doi.org/10.22331/q-2019-12-02-205},
  title = {Magic {S}tate {D}istillation: {N}ot as {C}ostly as {Y}ou {T}hink},
  author = {Litinski, Daniel},
  journal = {{Quantum}},
  issn = {2521-327X},
  publisher = {{Verein zur F{\"{o}}rderung des Open Access Publizierens in den Quantenwissenschaften}},
  volume = {3},
  pages = {205},
  month = dec,
  year = {2019}
}

@article{PRXQuantum.5.010337,
  title = {Partially Fault-Tolerant Quantum Computing Architecture with Error-Corrected Clifford Gates and Space-Time Efficient Analog Rotations},
  author = {Akahoshi, Yutaro and Maruyama, Kazunori and Oshima, Hirotaka and Sato, Shintaro and Fujii, Keisuke},
  journal = {PRX Quantum},
  volume = {5},
  issue = {1},
  pages = {010337},
  numpages = {21},
  year = {2024},
  month = {Mar},
  publisher = {American Physical Society},
  doi = {10.1103/PRXQuantum.5.010337},
  url = {https://link.aps.org/doi/10.1103/PRXQuantum.5.010337}
}

@misc{Thorvaldson2026,
  author = {Ian D. Thorvaldson and Jeffrey Marshall and Jack R. Craig and Samuel K. Gorman and Charles D. Hill and Michelle Y. Simmons},
  title = {The Magic Scroll: Leveraging biased noise to improve magic state cultivation in register-based architectures},
  year={2026},
  eprint={2608.09018v1},
  archivePrefix={arXiv},
  primaryClass={quant-ph}
}

@misc{Heinz2026,
  author = {Irina Heinz and Mira Sharma and Joris Kattemölle},
  title = {Exchange-only qubit stabilized by a single-spin qubit},
  year={2026},
  eprint={2608.14214},
  archivePrefix={arXiv},
  primaryClass={quant-ph}
}

\appendix

\section{Pulse sequence compilation tool}
\label{sec:appendix}

\begin{figure}[ht]
    \centering
    \includegraphics[width=.99\columnwidth]{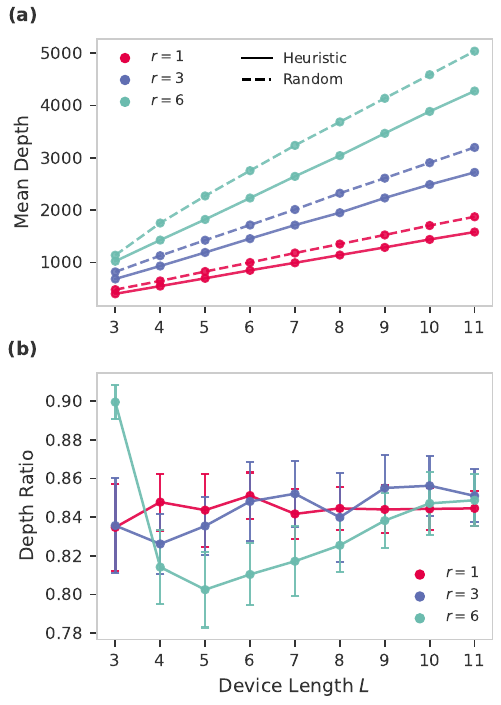}
    \caption{Circuit depth for restricted parallelization as different exclusion radii. (a) The heuristic critical-path algorithm (solid lines) is compared with the mean depth obtained from a randomized version of the same algorithm (dashed lines). (b) Depth ratio of the  heuristic vs. randomized algorithm, indicating a significant reduction in depth for all device sizes and exclusion radii.}
    \label{fig:heuristic_algorithm}
\end{figure}

To evaluate the operational feasibility of the architecture, we developed a stateful compilation tool built on the Google \texttt{Cirq} framework that automates the translation of logical EO qubit circuits into executable physical spin-qubit pulse sequences. The compilation process is based on a database containing preprocessed, parameterized pulse sequences derived from literature \cite{Chadwick2025,madzik_operating_2025, zeuch_efficient_2020}. Each entry in the database defines both the index-based pulse instructions—acting on physical chains of three or six spins—and the resulting spatial layout of the logical spin labels. Crucially, the database includes both non-permuting sequences and permuting sequences that change the internal arrangement of logical roles of spins inside each EO qubit. By tracking these transitions dynamically using an internal-state mapping, the compiler ensures that proper pulse sequences are read from the database at each stage of the circuit. Allowing spin positions to permute throughout the circuit allows for more hardware efficient gate implementations \cite{Chadwick2025}.

When executing multi-qubit operations, the tool extracts the localized physical chain of involved spins by taking into account the exact geometric layout of the participating EO qubits. This extraction accounts for the physical orientations of EO qubits on the chip architecture and automatically detects the underlying hardware connectivity. Single-qubit logical gates are mapped down to three-spin chains via an identical method.

After constructing the physical pulse circuit, the contained pulses are scheduled to analyze different levels of parallelization. For finite exclusion radii, pulse operation scheduling is achieved via a critical weight heuristic scheduling algorithm. The scheduler constructs a directed acyclic graph of the pulse dependencies, weights each node according to its dependencies, and prioritizes critical-path operations, i.e.~those pulses that have the most other pulses depend on it downstream in the circuit.

To assess the efficacy of the heuristic scheduling algorithm, we compare its resulting circuit depth against a version of the same algorithm that assigns random priorities when scheduling pulses. Fig.~\ref{fig:heuristic_algorithm}(a) illustrates the depth scaling of the heuristic approach (solid lines) relative to the mean randomized result (dashed lines) at exclusion radii $r\in\{1, 3, 6\}$. The depth ratio of the two methods is shown in Fig.~\ref{fig:heuristic_algorithm}(b), indicating that the heuristic algorithm significantly reduces the average depth by over 10\% for all exclusion radii. Error bars indicate the variance over different random scheduling runs.

\end{document}